\documentclass[twocolumn,aps,prb,showpacs,raggedbottom,nobalancelastpage,amssymb]{revtex4-1}

\usepackage{amssymb,amsmath,latexsym}
\usepackage{float}
\usepackage{dsfont}

\usepackage[usenames]{color}
\usepackage{soul}
\setulcolor{red}
\setstcolor{red}
\sethlcolor{yellow}

\newcommand{\ket}[1]{|#1\rangle}

\usepackage{graphicx} %Grafiken einbinden
\usepackage{subfigure} 
\usepackage{calc}
\usepackage{array}
\usepackage{url}
\usepackage{fancybox,amssymb,amsmath}
\usepackage{bbm} % fuer \mathbbm{1} - Einheitsmatrix
 \usepackage[title]{appendix}
\begin{document}
\title{Decoherence effects on weak value measurements in double quantum dots}

\author{Mark Thomas}

%\affiliation{Department of Condensed Matter Physics, The Weizmann Institute of Science, Rehovot 76100, Israel.}
%\affiliation{Dahlem Center for Complex Quantum Systems and Fachbereich Physik, Freie Universit\"at Berlin, 14195 Berlin, Germany}

\author{Alessandro Romito}

\affiliation{\mbox{Dahlem Center for Complex Quantum Systems and Fachbereich Physik, Freie Universit\"at Berlin, 14195 Berlin, Germany}}
%\affiliation{Dahlem Center for Complex Quantum Systems and Fachbereich Physik, Freie Universit\"at Berlin, 14195 Berlin, Germany}

\date{\today}
\begin{abstract}
We study the effect of decoherence on a weak value measurement in a
paradigm system consisting of a double quantum dot continuously
measured by a quantum point contact. Fluctuations of the parameters
controlling the dot state induce decoherence.
We find that, for measurements longer than the decoherence time, weak
values are always reduced within the range of the eigenvalues of the
measured observable. 
For measurements at shorter time scales, the measured weak value
strongly depends on the interplay between the decoherence dynamics of
the system  and the detector backaction.
In particular, depending on the
postselected state and the strength of the decoherence, a more frequent classical readout of the detector might lead to an enhancement of weak values.
\end{abstract}


\maketitle

\section{Introduction}

In quantum mechanics the measurement process is most simply described
as a probabilistic event through the projection postulate.~\cite{Neumann}
While it satisfactory describes several simple experimental
configurations, some measurement protocols, including
conditional quantum measurements, can lead to results that
cannot be interpreted in terms of classical probabilities, due to the
quantum correlations between measurements.
A striking evidence of that is provided by the so-called \emph{weak
  values} (WVs) obtained from the measurement scheme originally
developed by Aharonov, Albert, and Vaidman.~\cite{Aharonov:1988}
The WV measurement protocol consists of (i) initializing the system in
a certain state  $|\, \Psi\, \rangle$ (preselection), (ii) weakly
measuring an observable $\hat{A}$ of the system by coupling it to a detector, and
(iii) retaining the detector output only if the system is eventually
measured to be in a chosen final state $|\, \Phi\,
\rangle$ (postselection).  
The average signal monitored by the detector will then be proportional
to the real part of the so-called WV
$\label{eq:weakvalue1} \left.\vphantom{\langle
  \hat{A}\rangle}\right._{\Phi}\langle  \hat{A}
\rangle^{\textrm{weak}}_{\Psi} = \langle\, \Phi \,| \,\hat{A}\,
| \Psi\, \rangle   /   \langle \,\Phi\, |  \Psi \rangle$.

The most surprising property of WVs is that they can be complex or negative~\cite{Aharonov:1988,
  Hosoya-Shikano:2010} whereas a strong conventional measurement would lead to
positively definite values. %A weak measurement of a component of
                                %the spin of a spin-1/2 particle for
                                %instance can even take the value
                                %100~\cite{Aharonov:1988}.
After the original debate on the meaning and 
significance of weak
values,~\cite{Peres:1989,Aharonov-Reply:1989,Leggett:1989} they have 
proven to be a successful concept in addressing fundamental problems
and paradoxes of quantum mechanics,~\cite{Aharonov:2002, Lundeen:2009}
in accessing elusive quantities (e.g., the definition of the time
a particle spends under a potential barrier in a tunneling
process,~\cite{Steinberg:1995} the direct measurement of the wavefunction~\cite{Lundeed:2009}), in defining measurements in counterintuitive situations
(e.g., the simultaneous measurement of two non-commuting
observables~\cite{Hongduo:2008,Aharonov:2002}), as well as in
generalizing the definition of
measurement.~\cite{Jordan:2012}
By now a  number of experiments in quantum optics has reported
the experimental observation of WVs and its application to
quantum paradoxes.~\cite{Hosten:2008,Yokota:2009, Simon:2011,Dixon:2009}
% Joint weak measurements with an entangled photon pair have further
% been used to directly monitor a photonic version of Hardy's
% paradox~\cite{Yokota:2009}.
Recently, a series of interesting works has explored the potential of
WVs measurement protocols for precision measurements.
Weak-value-based measurement techniques have been successfully
employed in  quantum optics experiments to access tiny
effects~\cite{Hosten:2008} %cite{Wiseman:2002}
and detect ultrasmall (subnanometric)
displacements.~\cite{Dixon:2009,Simon:2011} 
Parallel research has introduced the idea of weak
values also in the context of solid state-state systems.~\cite{Romito-Gefen-Blanter:2008, Jordan-Williams:2008,
  Romito-Shpitalnik-Gefen:2008} 
Here, further works have shown that weak
values are related to the violation of classical inequalities in
current correlation measurements,~\cite{Bednorz-Belzig:2010}
and a WV measurement technique for ultrasensitive charge
detection has also been proposed.~\cite{Zilberberg:2011}
% \aleadd{decoherence is an issue, especially for solid state ones}
% Due to present day technology, many solid state systems are suitable
% to implement in quantum measurements~\cite{Korotkov:2001,
%   Zilberberg:2011}. %ultimately motivated by perspective quantum
%                     %computation.

% where conditional measurement in solid state systems has become the subject of increasing attraction. 

Due to the fact that WVs stem from quantum-mechanical
correlations between two measurements they are expected to be
particularly sensitive to decoherence.
The effect of decoherence is important for WV ultrasensitive measurements
where decoherence could suppress the amplification effect and become crucial
in possible solid-state implementations, where it is known that decoherence plays a significant
role in most systems. 
This is in fact the case for all the actual proposed implementations of WVs
in solid state
systems.~\cite{Jordan-Williams:2008,Romito-Gefen-Blanter:2008,Romito-Shpitalnik-Gefen:2008,Zilberberg:2011,Morello:2010}
%  and
% spin~\cite{Romito-Gefen-Blanter:2008} qubits in quantum dot, 
% to electronic Mach-Zehnder~\cite{Romito-Shpitalnik-Gefen:2008} and Aharonov-Bohms~\cite{Zilberberg:2011}, to nitrogen-vacancy centres~\cite{Morello:2010}.
So far, the effect of decoherence on
WVs has recently  been addressed at a formal level showing how
WVs are defined in a general open quantum
system,~\cite{Hosoya-Shikano:2010} while a quantitative evaluation of
the effects of decoherence in a specific system exists only for WVs of spin
qubits in a simple limit~\cite{Romito-Gefen-Blanter:2008, Romito:2009} and for correlated spin measurements of (unpolarized) electronic currents.~\cite{Lorenzo:2004}
Therefore,  a general characterization of the effects of decoherence within a weak
value measurement in an open quantum system is, in addition to its
theoretical significance, a relevant step in the direction of weak
value implementation in condensed matter systems.

In this work we precisely address this question.
We approach the problem by considering the effect of decoherence in a paradigm
system, namely a quantum point contact (QPC) sensing the charge in a nearby double
quantum dot.~\cite{Gurvitz:1997,Gurvitz:2008}
The model captures all the essential features of a continuous quantum
measurement, corresponding to the typical measurement schemes
of quantum states in nanoscale solid-state systems (which is the case for all the
above-mentioned proposals), and allows us to fully describe the
interplay between the detector backaction and the decoherence
process.

The key features of our analysis and the main results are as follows.
We describe the double quantum dot as a two-level system, $\ket{L}$,
$\ket{R}$,  corresponding to the electron being in the left or right dot, respectively.
In the system dynamics, we introduce fluctuations of the system's
parameter, for example the gate voltages, that suppress the quantum
mechanical  oscillations between these two states,
  at the decoherence time, $1/\gamma$.
The QPC detector, while distinguishing between the two system states,
affects the system dynamics on the scale of the backaction time,
$\tau_D$. As long as the measurement duration, $\tau$, is shorter
than $\tau_D$, the measurement is weak and can lead to WVs
upon proper postselection. The postselection, obtained, for instance by a second detector, is effectively
described as a projective measurement on a specific state, $\ket{\psi_f}$.

The quantity of interest in such a WV protocol is the detector signal conditional to a 
positive postselection, with a particular attention to the apparence of peculiar weak
values, that is, WVs which lie beyond the range of eigenvalues of the measured observable. By taking advantage of the Bayesian formalism, which allows us to
consider the correlations for single shot measurements, we obtain a
general expression for the WV in terms of the system state
only [cf. Eqs.~(\ref{eq:weak_current}) and (\ref{eq:perturbation_omega})].

We identify two different regimes  depending on whether the detector
readouts (i.e., relaxation processes) are slow or fast compared to the
duration of the measurement.
In both regimes, measurements longer than the decoherence time lead to
the WVs bounded by the eigenvalue spectrum.
In the former case, dubbed coherent detection, we show that the weak
value is exclusively determined by the system dynamics undergoing
decoherence --- cf. Eqs.~(\ref{eq:same_result}) and Figs.~\ref{fig3} and~\ref{fig4}.
In the latter case, named continuous readout, instead, even at time
scales $\tau \ll \tau_D$, the WVs are affected by the interplay of the 
detector and the decoherence dynamics [cf. Eqs.~(\ref{eq:n_multiple_readouts}( and (\ref{result-continuous}) and Figs.~\ref{fig5-6} and~\ref{fig5}]. In the coherent detection regime, the WV is sensitive to the average quantum-coherent correlation between measurement and postselection [Eq.~(\ref{eq:correlation})], vanishing for long-time measurements. The frequent projection in the continuous regime freezes the postselection, leading to a finite WV for long measurements. This difference reflects at shorter time scales, where a continuous detection can enhance the corresponding WV obtained for a coherent detection. In particular, depending on the various parameters, for example orientation of
the postselection, it could give rise to WVs beyond the range of
eigenvalues, whereas a coherent detection would not (cf. Fig.~\ref{fig5}).

%We show that while the decoherence generally reduces the WV within the range of standard values, the specifics of the decay strongly depend on the detector dynamics, and, in some regimes, induce an enhancement of the WV.

% We are specifically interested in weak measurements in solid state
% systems, which typically (as for all the above-mentioned proposals),
% involve continuos quantum measurements. 
% Rather than a general formal expression for WVs\alecom{check
%   again this sentence, it is important} \markcom{Eq.~(\ref{eq:permutations}) is a general result of a frequent readout}, we approach
% the problem by addressing the effect of decoherence in a paradigm
% system that captures the main features of a continuous detection,
% namely a quantum point contact (QPC) sensing the charge in a nearby double
% quantum dot.~\cite{Gurvitz:1997, Reilly:2007, Küng:2010} \alecom{cite gurvitz and first exp papers on that} \markcom{first papers?}. 

The paper is primarily separated into four
parts. In Sec.~\ref{sec:model} we present the model and its
description in terms of the Bayesian formalism. 
Hereafter, Sec.~\ref{sec:weakvalue} presents a general expression for the weak
value in terms of the density matrix of the combined  qubit-detector system. 
In Sec.~\ref{sec:section4} we discuss the system dynamics of the two regimes of ``coherent detection''~(Sec.~\ref{sec:single_read-out}) and continuous readout~(Sec.~\ref{sec:continuous_read-out}).
In the former regime,
the detector and system dynamics are decoupled from each other in the
weak measurement regime; in the latter regime, we show that 
the detector induced decoherence and the intrinsic system
decoherence act together in affecting the measurement outcome. Section~\ref{sec:results} summarizes our results.

\section{The Model}\label{sec:model}

\begin{figure}[t]
\includegraphics[width=0.4\textwidth]{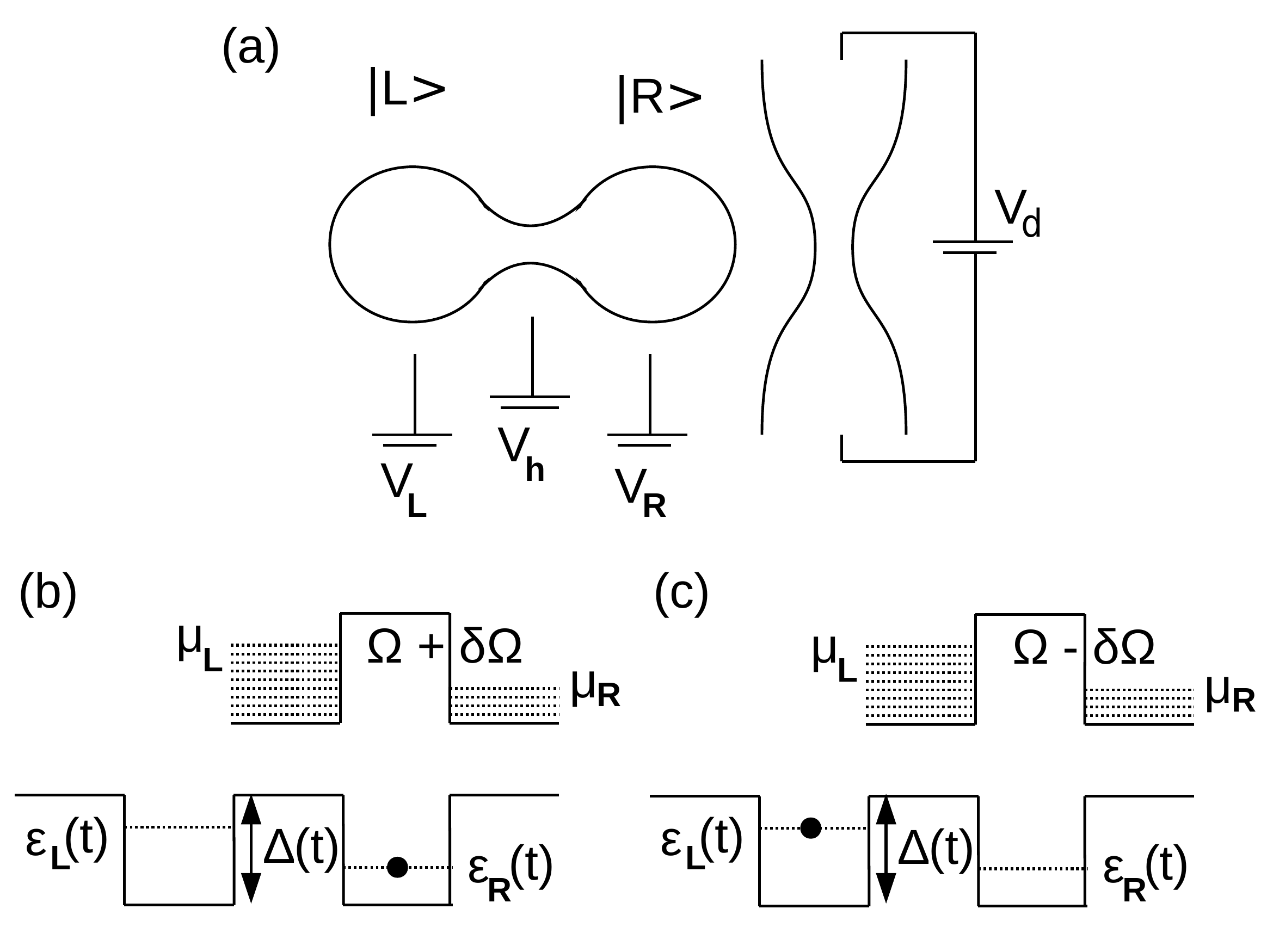}
	\caption{(a) Scheme of a double quantum dot  with  a nearby
          QPC  measuring its charge states, labeled as $\ket{L}$, $\ket{R}$. 
(b), (c) Scheme of the detection mechanism. The conductance of the detector is
          sensitive to the qubit's occupancy: The tunneling amplitude in
the  QPC is $\Omega + \delta \Omega$ when the right dot is occupied
(b)  and $\Omega - \delta \Omega$ when the left dot is occupied (c).} 
	\label{fig:system}
\end{figure}

The system under study is illustrated in
Fig.~\ref{fig:system}. It consists of a double quantum dot with
an adjacent QPC, which serves as a detector continuously measuring the
charge state in the double dot.

We consider the case where, due to charging
effects, the double dot can host only one electron in two orbital
levels, $\ket{L}$ and $\ket{R}$ corresponding to the lowest orbital
level for the left and right dot, respectively.  
In this case the double dot can be thought as a charge qubit.
Throughout the paper we refer to the ``dot plus QPC'' as the
combined qubit-detector system which is then described by the Hamiltonian
\begin{equation}\label{eq:Hamiltonian}
H(t) = H_0(t)  + H_{\textrm{int}} + \hat{H},
\end{equation}
where $H_0(t)$ denotes the Hamiltonian of the double dot,
$H_{\textrm{int}}$ characterizes the system-detector interaction and
$\hat{H}$ describes the Hamiltonian of the detector. 
Specifically
\begin{align}
& H_0(t) =  \vphantom{\frac{1}{1}} \epsilon(t)\, \sigma_z + \Delta(t)\, \sigma_x  \vphantom{\left( \sum_r\right)}\\
& \hat{H} +H_{\textrm{int}} = \frac{1}{2}  \left( \Delta
  \hat{H}^{+} + \Delta
  \hat{H}^{-} \otimes \sigma_z \right) 
\label{main}
\end{align}
where
\begin{align}
\Delta \hat{H}^{+} =& \sum_r E_r a_r^{\dag} a_r + \sum_l E_l
a_l^{\dag} a_l +\sum_{r,l} \left( \Omega_{lr} a_l^{\dag} a_r + h.c. \right) \\
\Delta \hat{H}^{-} =& 2 \sum_{lr}  \left( \delta \Omega_{lr}
  a_l^{\dag} a_r + h.c. \right)
\label{main1}
\end{align}
Here $\sigma_z = |\,\textrm{R}\,\rangle \langle\, \textrm{R}\,| -
|\,\textrm{L}\,\rangle\langle\, \textrm{L}\,|$, $\sigma_x =
|\,\textrm{R}\,\rangle\langle\, \textrm{L}\,| +
|\,\textrm{L}\,\rangle\langle\, \textrm{R}\,|$ and
$|\,\textrm{R}\,\rangle$ , $|\,\textrm{L}\,\rangle$ are, without loss
of  generality, the eigenstates of the operator measured by the
QPC.~\footnote{In the general case where the eigenstates of the
  operator measured by the QPC do not coincide with   $|L>$, $| R >$,
  $H_0$ written in such an eigenstates basis takes the same form with
  renormalized coefficients $\epsilon$, $\Delta$.} 
%foot1 DO NOT USE %THE COMMAND RANGLE OR LANGLE
%%HERE!!!!!!!!!!!!!!!!!!!!!!!!!!!!!!!!    
The on-site energy difference, $2 \epsilon(t)$, and the tunneling
between such states,  $\Delta(t)$, are controlled by the externally
applied voltage  biases, $V_L$,$V_R$, $V_h$.
In Eqs.~(\ref{main}) and (\ref{main1}), the operators in the QPC space are indicated by a $\hat{\,}$,
and we do so henceforth.
The creation and annihilation operator in the two leads of the QPC are
denoted by $a^{\dagger}_i$ and $a_i$ respectively, where
$i=l,r$ refer to the left and the right reservoir,
respectively. 
$E_i$ characterizes the energy of the reservoir states, which are
maintained at the corresponding Fermi energies $\mu_L =\mu_R+eV_d> \mu_R$, and
$\Omega_{l,r} \pm \delta \Omega_{l,r}$  denotes the tunneling
amplitude between the reservoirs when $|\,\textrm{R}\,\rangle$ or
$|\,\textrm{L}\,\rangle$ is occupied.

This model has been extensively studied in the
literature.~\cite{Gurvitz:1997,Gurvitz:1998,Korotkov:1999,Korotkov:2001}
The current through the detector directly measures the ``position''
of the electron in the dot.
Under the assumptions of uniform tunneling matrix elements, $\delta\Omega_{l,r}
\equiv \delta\Omega$, $\Omega_{l,r} \equiv \Omega$, and density
of states in the QPC's reservoirs, $\rho_L$, $\rho_R$, the average
current for an electron being in the left or right
dot, reads $
 \langle I \rangle_{\pm} = e D/2 \cdot ( 1 \pm
 \delta\Omega/\Omega)^2 $, where $D \equiv T V_d/(2 \pi) =
 2\pi\rho_l\rho_r\Omega^2V_d$. Throughout the work we set $\hbar = 1$.
The QPC's shot noise power $S_I =e I (1-T)$ sets the time scale 
$\tau_D \sim S_I/(I_+-I_-)^2\approx 1 /D (\Omega/\delta \Omega)^2
(1-T)$  needed to distinguish the detector's signal form the
background noise. 
The weak measurement regime is then identified by measurements of
duration $\tau \ll \tau_D$, which can be controlled in principle by
tuning the system-detector coupling, the duration of the measurement,
or the voltage bias across the QPC.  
The QPC is effectively a detector at finite $V_d$ where it leads to a finite signal $\langle I \rangle \propto V_d$. In particular, we assume $V_d \gg k_B T$ (the temperature $T$ is the smallest energy scale throughout the paper) and $V_d \gg \rho \Omega^2$ as clarified below. In this regime we neglect the extra decoherence effect due to the detector equilibrium backaction, for instance orthogonality catastrophe dephasing.~\cite{Aleiner:1997,Gefen:2012}
%It is important to notice that, in the present model of the QPC, we
%assume that the equilibrium state of the reservoirs is not affected by
%$\Omega_{lr}$. 
%In other words the backaction of the detector is only due to
%out-of-equilibrium effects ($V_d=0$), while we are neglecting possible
%equilibrium backactions, e.g. orthogonality catastrophe dephasing.~\cite{Aleiner:1997,Gefen:2012}

\begin{figure}[t]
	\begin{centering}		
	 \includegraphics[width=0.45\textwidth]{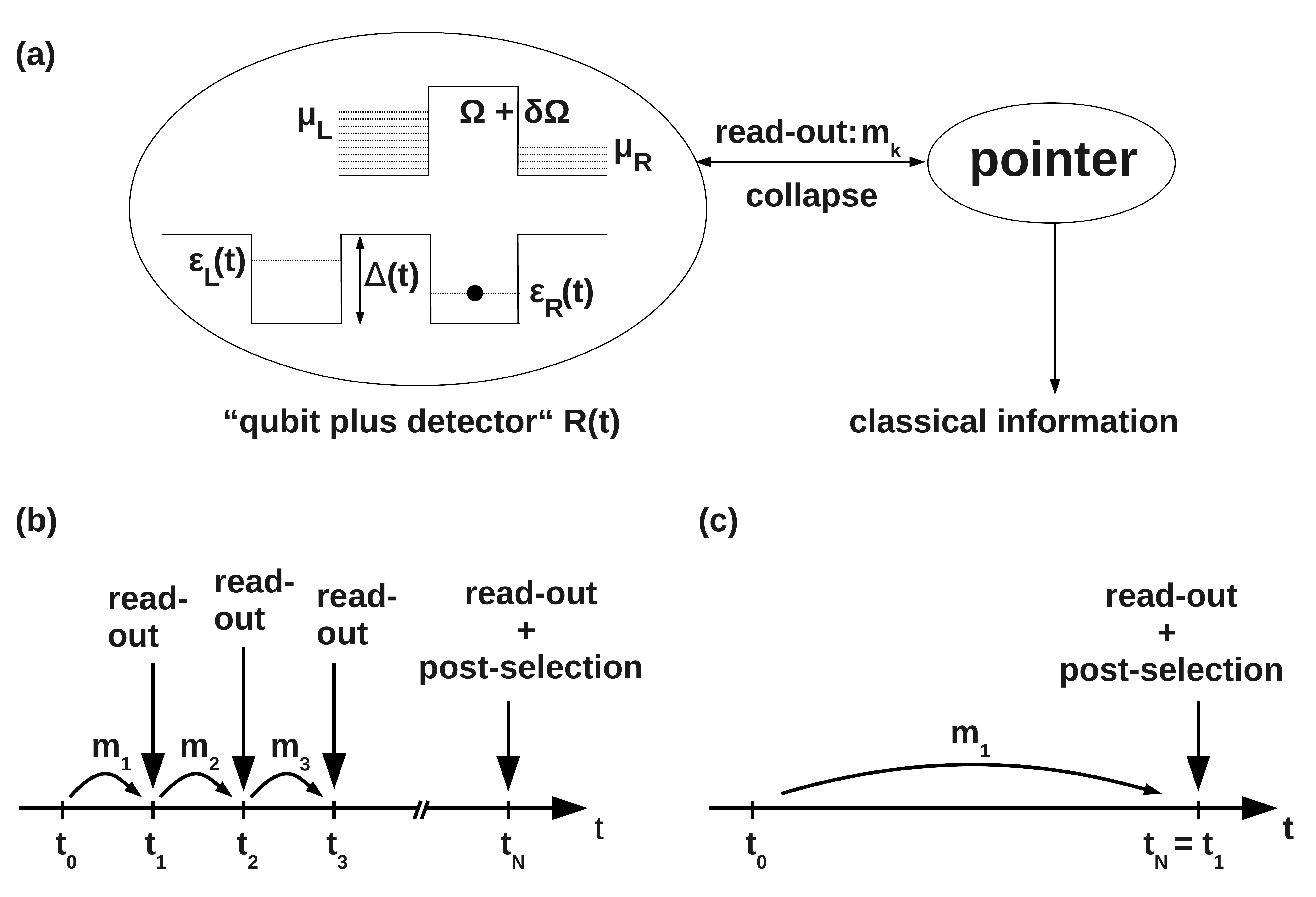}
	\end{centering}
	\caption{(a) Qubit plus detector interaction with a classical
          pointer reading out the number $m_k$ of electrons
           having penetrated to the detector's right reservoir at
           certain discrete times $t=t_k$. Illustration of the continuous
           readout (b) and coherent (c) measurement regimes,
           respectively.}
	\label{fig:system2}
\end{figure}

An important aspect of the model is that the Hamiltonian in
Eq.~(\ref{eq:Hamiltonian}) describes the qubit and  the detector as a
closed quantum system.  
However, the QPC is continuously converting the information about the
state of the system in a \emph{classical} --- macroscopic ---
information output which is the current. 
In other words, while Eq.~(\ref{eq:Hamiltonian}) will evolve the
detector to a \emph{coherent} superposition of states with different
charges in the reservoirs, the classical knowledge of the current
would correspond to a well-defined number of electrons in each
reservoir. 
A solution of this problem, as pointed out in
Ref.~\onlinecite{Korotkov:2001}, consists of introducing a macroscopic
pointer which interacts with the detector. The pointer
provides in fact an effective description of the various
relaxation processes confining the QPC electrons in one of the
two QPC reservoirs.
% It is important to note that the external pointer interacts
% \textit{classically} with the full ``dots and detector" system since
% it reads a macroscopic detector current whereas the detector itself is
% described \textit{quantum mechanically}.
The pointer can be modeled to interact instantaneously at certain
times, $t_1, t_2 \ldots t_k \ldots t_N$ with $t_0 =0$ and $t_N = \tau$,
and  reads out the change of the number of electrons in the right
reservoir, $\mu_R$,  as schematically depicted in
Fig.~\ref{fig:system2}. At any time $t=t_k$ the pointer reads the number of electrons,
$m_k$, transmitted to the collector within the time interval
$\Delta t_k = t_{k+1} - t_k$, and collapses the qubit-detector system onto a
corresponding state depending on the measured value $m_k$.
Note that the introduction of the pointer results in a new time scale
$\Delta t_k$ in the problem which is a free parameter in our model. We
 discuss in the Results section the different regimes corresponding to the
relation of this time scale to other time scales in the problem.
% Note that the coupling between the system and the detector is controlled by $\delta \Omega$ meaning that a decoupled decoupled detector has a non-vanishing current $\langle I\rangle = e D/ (2\hbar)$.

The model allows us to include decoherence by considering
fluctuations, which naturally arise in the system, of the
external parameters, namely the voltage biases $V_L$,$V_R$, $V_h$. In
turn they lead to fluctuations of the dots' parameters, $
 \epsilon(t) = \epsilon_0 + \xi(t)\quad$ and $\Delta(t) = \Delta_0 + \eta(t) $
around their average values $\epsilon_0$ and $\Delta_0$. Since the
explicit relation between $V_h$, $V_L$, $V_R$ and $\epsilon_0$,
$\Delta_0$ depends on microscopic details, we effectively assume here
a  general linear relation and the effect of decoherence is therefore
described by replacing the dot's Hamiltonian with $ H_0(t) \rightarrow
H_0 + H_{\xi}(t)$, with
\begin{align}
 & H_0 = \epsilon_0 \,\sigma_z + \Delta_0\,\sigma_x  = \omega
 \mathbf{m} \cdot \mathbf{\sigma}, \\
& H_{\xi} (t) = \sum_i\, \xi_i(t) \, \mathbf{k}_i\cdot \mathbf{\sigma}.
\end{align}
Here $\omega=\sqrt{\epsilon_0^2+\Delta_0^2}$ is the oscillation
frequency of the system and $\mathbf{m}$ defines 
the corresponding eigenstates.
To this regard, the index $i$ labels the different \emph{independent}
decoherence sources, ideally corresponding to the independent voltage
sources. For each of them  $\mathbf{k}_i$ indicates the direction of the fluctuations with $|\mathbf{k}_i| = 1$ and $\xi_i(t)$ is assumed to be a Gaussian white noise, that is,
\begin{equation}\label{eq:white_noise}
 \left\langle \, \xi_i(s)\,\right\rangle_{\xi} = 0, \quad  \left\langle \, \xi_i(s_1)\,\xi_j(s_2) \right\rangle_{\xi} =  \gamma_i\,\delta(s_1 - s_2)\,\delta_{ij}
\end{equation}
where $\gamma_i$ describes the strength of the correlation
function. For the sake of simplicity we present in the following our general results
for the case of a single decoherence source. The results in the case
of several noise sources are a straightforward extension and are
discussed at the end.

Finally, we can include in the model the description of
postselection, as required by the WV measurement protocol.
As pointed out in Ref.~\onlinecite{Romito-Gefen-Blanter:2008}, 
a second QPC strongly measuring the charge on the dot at any time after
the weak measurement can effectively realize a postselection in any
given qubit state, $\ket{\Psi_f}$. 
% ~\footnote{Another possibility for realising the
%   postselection is that of a pulsed measurement by only using one
%   detector. In this case, a first pulse of duration $\tau$ would
%   perform a weak measurement and a subsequent one would relate would
%   relate the strong measurement.} %no footnote!!!
Without loss of generality we therefore consider the situation where
the postselection takes place immediately after the weak measurement.
Within our model the postselection into the state $\ket{\Psi_f}$ is
described by the action of the corresponding  projection operator
$\Pi_f$ acting at the end of the weak measurement.

\section{General Expression for the Weak Value in the Presence of Decoherence}\label{sec:weakvalue}

The WV protocol we are interested in consists of preparing the
double dot in a given initial state $\ket{\Psi(t_0)}$ at time $t_0$,
making the system interact with the detector for a time $\tau$, and
 finally strongly measuring it  in the postselected state in $\ket{\Psi_f}$.
The quantity of interest is the WV of the electron's
occupancy in the double dot, that is, the value of $ \sigma_z(\tau)$
conditional to a positive postselected outcome, which we denote by $ _{\Pi_{\textrm{f}}}\langle \sigma_{z}(\tau)
\rangle_{\Psi(t_0)}$.
In fact, such a quantity is inferred from the postselected output of
the detector, that is, the average current in the QPC  
conditional to the postselection $_{\Pi_{\textrm{f}}}\langle I(\tau)
\rangle_{\Psi(t_0)}=  e/\tau \,
_{\Pi_{\textrm{f}}}\langle n(\tau) \rangle_{\Psi(t_0)} $ through 
\begin{equation}
 _{\Pi_{\textrm{f}}}\langle \sigma_{z}(\tau)
\rangle_{\Psi(t_0)} = \frac{_{\Pi_{\textrm{f}}}\langle I(\tau)
\rangle_{\Psi(t_0)} -eD/2h}{ I_+-I_-} \,.
\end{equation}
 The average conditional (postselected) value of the number of
 electrons, $_{\Pi_{\textrm{f}}}\langle n(\tau) \rangle_{\Psi(t_0)}$,
 having  passed through  the QPC during the measurement time
 $\tau$ is  
\begin{equation}\label{eq:weak_n}
\left.\vphantom{\langle \Omega
    \rangle}\right._{\Pi_{\textrm{f}}}\langle n(\tau)
\rangle_{\Psi(t_0)} =  \sum_{\substack{m}}  m\, \left. P(m \,\right|\,
\Pi_f ) \, .
\end{equation}
In Eq.~(\ref{eq:weak_n}) $m$ indicates the total number of electrons
having reached the collector and  $P(m \,|\, \Pi_f )$ is the
conditional probability that $m$ electrons have  been transmitted
through the QPC given that the qubit  is finally found to be in a state
represented by the projection operator $\Pi_f$. 
Note that Eq.~\eqref{eq:weak_n} is valid for any strength of the
measurement.
 We keep our analysis valid for a general measurement strength until specified differently.

The conditional probabilities in Eq. (\ref{eq:weak_n}), can be
directly expressed in terms of the total density matrix $R(t)$ of
the qubit-detector system. 
Following the formalism of Gurvitz~\cite{Gurvitz:1997} and
Korotkov,~\cite{Korotkov:2001}  a pure state of the qubit-detector system is described by a wave function $	\left| \, \Psi(t)\,\right\rangle = (  | \, \Psi^{\uparrow}(t)\,\rangle, | \, \Psi^{\downarrow}(t)\,\rangle )$,
where $\sigma = \uparrow, \downarrow$ labels the eigenstates of $\sigma_z$ and $| \Psi^{\sigma}(t) \rangle$ is a  many-body state of the QPC,
\begin{align}\label{eq:many-body_states}
	\left| \, \Psi^{\sigma}(t)\,\right\rangle &= \left( b^{(0)\,\sigma}(t) + \sum_{\substack{ l \leq \mu_l; r > \mu_r}}  b^{(1)\,\sigma}_{lr}(t)\,a_r^{\dagger}a_l \right. \notag\\
	&\left. + \sum_{\substack{ l,l' \leq \mu_l; r,r' > \mu_r}}  b^{(2)\,\sigma}_{ll'rr'}(t)\,a_r^{\dagger}a_{r'}^{\dagger}a_l a_{l'} + \cdots
	\right) \, | 0 \rangle \, ,
\end{align}
and $|b^{(i)}(t)|^2$ describes the probability of finding the entire
qubit-detector system in the corresponding  state described by the creation and
annihilation operators with $l$, $r$  labeling the single particle
states in the left, right reservoirs, respectively. 
The corresponding qubit-detector density matrix $R(t)$  has components 
\begin{equation}\label{eq:full_rho}
	R_{\sigma \sigma'}(t) = \left( \begin{array}{ccccc} 
	R^{(0,0)}_{\sigma \sigma'}(t) & R^{(1,0)}_{\sigma \sigma'}(t) & \ldots &  &  \\
	R^{(0,1)}_{\sigma \sigma'}(t) & R^{(1,1)}_{\sigma \sigma'}(t) & \ddots &   &                              \\
	     \vdots                         &    \vdots    &    & R^{(m,n)}_{\sigma \sigma'}(t) & \ldots
	\end{array} \right) \, .
\end{equation}
Here each entry $R^{(m,n)}_{\sigma \sigma'}(t)$ is a matrix whose
dimensions are given by the  infinitely many states labeled by $l, r,
l', r', l'', r'' \ldots l^{(m)}, r^{(m)}$  and $l, r, l', r', l'', r''
\ldots l^{(n)}, r^{(n)}$. 
Each of the entries $b_{l_1\ldots l_m r_1\ldots r_m}^{(m)\,\sigma}(t)
\, \overline{b}_{l_1\ldots l_n r_1\ldots r_n}^{(n)\,\sigma'}(t)$ (the
$\overline{\cdot}$ indicates the  complex conjugate) in
$R^{(m,n)}_{\sigma \sigma'}(t)$  characterizes the coherences 
between all the possible states with $m$ and  $n$ electrons 
detected in the collector of the QPC at time $t$. 
In particular, the trace of each diagonal matrix $R^{(m,m)}_{\sigma
  \sigma'}(t)$ identifies  the probability that exactly $m$ electrons
have passed  through the detector until time~$t$, 
namely
\begin{equation}\label{eq:Pm}
 P(m) = \textrm{tr}_{\textrm{sys}}\left\{ \mathcal{R}^{(m)}(t)
 \right\} = \textrm{tr}_{\textrm{sys}} \left\{
   \textrm{tr}_{\textrm{det}} \left\{ R^{(m,m)}(t) \right\} \right\} ,
\end{equation}
where we introduced the quantity $\mathcal{R}_{\sigma \sigma'}^{(m)}(t) = \textrm{tr}_{\textrm{det}}\{ R_{\sigma \sigma'}^{(m,m)} \}$. Also the reduced density matrix of the dot, $ \rho(t) = \textrm{tr}_{\textrm{det}}\left\{ R(t)
 \right\}$, can be written as 
\begin{align}
 \rho(t) = \sum_{\substack{m}} \mathcal{R}^{(m)} = \sum_{\substack{m}} P(m) \cdot
 \rho^{(m)}(t) \, ,
\label{eq:partial_trace} 
\end{align}
where 
$ \rho^{(m)}(t) = \mathcal{R}^{(m)}(t) / (\textrm{tr}\left\{
  \mathcal{R}^{(m)}(t) \right\} )$
describes the state of the qubit where $m$ electrons have reached the collector.

Besides the inherent quantum-mechanical fluctuations, the stochastic
parameter $\xi(t)$ assumes different values at each replica of the
experiment according to its probability distribution. 
In order to properly take into account the average over fluctuations,
we can first rewrite the conditional probability
Eq.~(\ref{eq:weak_n})  using Bayes' theorem as $P(m | \Pi_f) = P(m) P(\Pi_f | m)/\sum_m P(\Pi_f)$.
The WV of each run of the
experiment is now weighted with the probability of 
a successful postselection in the corresponding run of the
experiment, which also depends on the specific noise realization.
This means that the average over the fluctuations $\xi(t)$ in the weak
value is properly taken into account by separately averaging over $\xi$ both  the
conditional average value of $m$ and the postselection probability.~\cite{Romito:2009} 
This leads to
\begin{equation}
 \left.\vphantom{\langle \Omega
     \rangle}\right._{\Pi_{\textrm{f}}}\langle n(\tau)
 \rangle_{\Psi(t_0)} =  \frac{\left\langle \sum_{\substack{m}} m \,
     \left. P(\Pi_f \,\right| ( m | \xi ) ) \,  P( m | \xi )
   \right\rangle_{\xi}}{\left\langle  \sum_{\substack{m}} \left.
       P(\Pi_f \,\right| ( m | \xi ) ) \, P( m | \xi )
   \right\rangle_{\xi}}  \vphantom{\frac{\frac{\Omega}{\Omega}}{\frac{\Omega}{\Omega}}}.
\end{equation}
Identifying the emerging probabilities in terms of the matrix $\mathcal{R}^{(m)}$ similar to Eq.~(\ref{eq:Pm}) yields the conditional current
\begin{equation}\label{eq:weak_current}
\left.\vphantom{\langle \Omega
    \rangle}\right._{\Pi_{\textrm{f}}}\langle I(\tau)
\rangle_{\Psi(t_0)} =  \frac{e}{\tau}  \,\frac{ \sum_{\substack{m}} m
  \, \textrm{tr}\left\{ \Pi_f \cdot \langle \mathcal{R}^{(m)}(\tau)\rangle_{\xi}
  \right\}  }{ \sum_{\substack{m}}  \textrm{tr}\left\{ \Pi_f  \cdot
    \langle  \mathcal{R}^{(m)}(\tau)\rangle_{\xi} \right\}  }.
\end{equation}
As already pointed out the weak measurement regime is obtained when
  $(\delta\Omega / \Omega)^2  \tau \ll  1/D (1-D/V_d)$.
The WV can be identified from the coefficients in the expansion
 \begin{equation}\label{eq:perturbation_omega}
 \left.\vphantom{\langle \Omega
     \rangle}\right._{\Pi_{\textrm{f}}}\langle I(\tau)
 \rangle_{\Psi(t_0)} = \langle I(\tau) \rangle +
 \left.\vphantom{\langle \Omega
     \rangle}\right._{\Pi_{\textrm{f}}}\langle I(\tau)
 \rangle^{\textrm{weak}}_{\Psi(t_0)} \cdot \frac{\delta\Omega}{\Omega}
 + \mathcal{O}\left(\frac{\delta\Omega^2}{\Omega^2} \right).
\end{equation}
We discuss the validity of this expansion later. From the
definition of the weak measurement regime, we expect to be sensitive
to coherence effects for time scales $\tau \ll 1/D (1-D/V_d)
(\Omega / \delta \Omega)^2$.
The WV is completely determined by the knowledge of the
probabilities $P(m)$  and the conditional reduced dot's density matrix
$\langle \rho^{(m)}(\tau) \rangle_{\xi}$.  
Further analysis now focuses on the evaluation of $\langle
\mathcal{R}^{(m)}(\tau)\rangle_{\xi}$.

\section{System-detector dynamics in presence of decoherence}\label{sec:section4}
\subsection{Coherent detection}\label{sec:single_read-out}

First we consider the case where a single readout by the external
pointer takes place at the end of the measurement process, that is, at
$t=\tau$, immediately followed by the postselection. 
In this case the coherent evolution of the system and detector is not
disturbed by the pointer, and we thus name it coherent detection.
 
The WV is fully determined by the knowledge of the averaged
matrices $\langle \mathcal{R}^{(m)}(\tau)\rangle_{\xi}$.
We can derive a differential equation for $\langle
\mathcal{R}^{(m)}(\tau)\rangle_{\xi}$ starting from the von Neumann
equation for the qubit-detector system, $i  \partial_t R(t) =
\left[ H(t), R(t)\right]$. 
After inserting the Hamiltonian in Eq.~\eqref{eq:Hamiltonian} with the
specific choice of a single fluctuation source,  that is, $\epsilon(t) =
\epsilon_0 + k_z \cdot \xi(t)$, $\Delta(t) = \Delta_0 + k_x \cdot
\xi(t)$, one obtains for the qubit-detector evolution
\begin{equation}\label{eq:evolution_non-average}
\partial_t  R(t) =
-i\,\left[\,M,\, R(t)\,\right]
-i\,\xi(t)\,\left[ \,N, \,R(t)\, \right]
\end{equation}
where $\partial_t = \partial / \partial t$. Here,
\begin{align}
 M &= \frac{1}{2}\, \Delta \hat{H}^{+} + \Delta \cdot \sigma_x +
 \frac{1}{2} \left( \Delta \hat{H}^{-} +  2 \,\epsilon_0 \right)
 \otimes \sigma_z \,, \notag\\
N &= \vphantom{\frac{1}{2}} \mathbf{\sigma} \cdot \mathbf{k}\,. 
\end{align}
Introducing an interaction picture with respect to $M$, which transforms the arbitrary operator $A$ as $A_I(t) = e^{-iMt} \cdot A \cdot e^{iMt}$, transfers Eq.~(\ref{eq:evolution_non-average}) into $\partial_t  R_I(t) =
-i\,\xi(t)\,\left[ \,N_I(t), \,R_I(t)\, \right]$. Iteratively solving this equation, one
obtains after taking the average with respect to $\xi(t)$
\begin{align}
 \left\langle R_I(t)\right\rangle_{\xi} &= \left\langle R(0)\right\rangle_{\xi} - i \int_0^t \textrm{d}s_1 \langle \xi(s_1) \rangle_{\xi} \, [N_I(s_1),\left\langle R_I(0)\right\rangle_{\xi}] \notag\\
 & + (-i)^2 \int_0^t \textrm{d}s_1 \int_0^{s_1} \textrm{d}s_2 \langle \xi(s_1) \xi(s_2) \rangle_{\xi} \notag\\
 &\quad \cdot [N_I(s_1),[N_I(s_2),\left\langle R_I(0)\right\rangle_{\xi}]] \notag\\
 & + (-i)^3 \int_0^t \textrm{d}s_1 \int_0^{s_1} \textrm{d}s_2 \int_0^{s_2} \textrm{d}s_3 \langle \xi(s_1) \xi(s_2) \xi(s_3) \rangle_{\xi} \notag\\
 &\quad \cdot [N_I(s_1),[N_I(s_2),[N_I(s_3),\left\langle R_I(0)\right\rangle_{\xi}]]] \notag\\
 & + \ldots
\end{align}
Due to the $\delta$-like time correlations in
Eq.~(\ref{eq:white_noise}),  the average can be performed exactly
order by order. The so
obtained integral equation for $R(t)$ is more conveniently written in
a differential form as 
\begin{equation}\label{eq:average_neumann}
\partial_t \left\langle
  R(t)\right\rangle_{\xi} =  -\,\frac{\gamma}{2}\, \left[\,
  N,\, \left[ N,\, \left\langle  R(t)\right\rangle_{\xi} \right]
\right]  -i\,\left[ \,M, \left\langle R(t)\right\rangle_{\xi}\,\right]
\end{equation}
The density matrix is expressed as a linear combination of
Pauli-matrices of the dot's space
\begin{equation}\label{eq:rho_and_v_j}
\left\langle  R(t)\right\rangle_{\xi} = \frac{1}{2}\,\sum_{j=0}^{3} \hat{v}_{j}(t) \cdot \sigma_j.
\end{equation}
where each $\hat{v}_{j}(t)$ is a matrix in the QPC Hilbert
space with $\textrm{tr} \left\{\hat{v}_{0}(t) \right\} \equiv
1$. 
Substituting Eq.~\eqref{eq:rho_and_v_j} into the averaged von Neumann
equation~\eqref{eq:average_neumann}  yields a set of differential equations
for $\hat{v}_{j}(t)$  which fully describes the averaged
evolution of the  qubit-detector system:
\begin{align}
 \dot{\hat{v}}_{0} &=  -\frac{i}{2}\,\left[ \Delta \hat{H}^{-}, \hat{v}_{z}(t) \right] -\frac{i}{2}\,\left[ \Delta \hat{H}^{+}, \hat{v}_{0}(t) \right]    \notag\\
\dot{\hat{v}}_{x} &= 2\gamma \left( \vphantom{\int} (k_x^2-1) \hat{v}_{x} +k_x k_y \hat{v}_{y} + k_x k_z \hat{v}_{z} \right)  -2\epsilon_0 \hat{v}_{y} \notag \\
&\quad- \frac{i}{2}\,\left[ \Delta \hat{H}^{+}, \hat{v}_{x} \right]  - \frac{1}{2}\,\left\{ \Delta \hat{H}^{-}, \hat{v}_{y} \right\}    \notag\\
\dot{\hat{v}}_{y} &= 2\gamma \left( \vphantom{\int} k_x k_y \hat{v}_{x} + (k_y^2-1) \hat{v}_{y} + k_y k_z \hat{v}_{z} \right) +2\epsilon_0 \hat{v}_{y}   \notag\\
& \quad  -2\Delta\hat{v}_{z} -\frac{i}{2}\,\left[ \Delta \hat{H}^{+}, \hat{v}_{y} \right]  +\frac{1}{2}\,\left\{ \Delta \hat{H}^{-}, \hat{v}_{y} \right\}    \notag\\
 \dot{\hat{v}}_{z} &= 2\gamma  \left( \vphantom{\int} k_x k_z \hat{v}_{x} + k_y k_z \hat{v}_{y} + (k_z^2-1) \hat{v}_{z} \right) + 2\Delta\hat{v}_{y}\notag\\
&\quad -\frac{i}{2}\,\left[ \Delta \hat{H}^{-}, \hat{v}_{0} \right]  -\frac{i}{2}\,\left[ \Delta \hat{H}^{+}, \hat{v}_{z} \right] \label{eq:ODEforv_z}.
\end{align}
Equations.~(\ref{eq:ODEforv_z}) describe a set of infinitely
many coupled differential equations.
They are a generalization to a density matrix of the results obtained
in Ref.~\onlinecite{Gurvitz:1997} for the simple case of a pure state. Note
that such a generalization is essential in our case to properly treat
fluctuations of $\xi$. One may further note that the fluctuations treated here differ from Ref.~\onlinecite{Korotkov:2001}, where the system's evolution for a given stochastic measurement output $I(t)$ is studied.  
Considering the matrix elements of $R(t)$ between two sectors $(m,n)$
with $m$ and $n$ electrons having passed through the QPC, we
trace out the QPC degrees of freedom to obtain a differential
equation for 
\begin{equation}
\langle \mathcal{R}^{(m)}(t) \rangle_{\xi} = \frac{1}{2} \,
\left(\mathbf{v}^{(m)}(t) \cdot \mathbf{\sigma} \right) \, ,
 \end{equation}
 where $v_j^{(m)}(t) = \textrm{tr}_{\textrm{det}}
\{\hat{v}_{j}^{(m,m)}(t) \}$. The details of the calculation are presented in
appendix~\ref{app:averaged_ODE}. Here we report the resulting differential equation, which 
can be recast as a differential equation for $v_j^{(m)}(t)$, that is,
\begin{equation}\label{eq:ODE}
 \partial_t \, \mathbf{v}^{(n)}(t) = \left(G_0 +
   G_{\mathbf{k}}+ G_1 \right) \cdot  \mathbf{v}^{(n)}(t) +  G_2 \cdot
 \mathbf{v}^{(n-1)}(t) \, ,
\end{equation}
with
\begin{equation} \notag
 G_0 = \left( \begin{array}{cccc}      
 \,\, 0 \quad\quad & 0 & 0 & 0 \\
 \,\, 0 \quad\quad & 0 & -2\,\epsilon_0 & 0 \\
 \,\, 0 \quad\quad & 2\,\epsilon_0 & 0 & -\,2\,\Delta_0 \\
 \,\, 0 \quad\quad & 0 & 2\,\Delta_0 & 0 
 \end{array}\right) ,
\end{equation}
\begin{equation}\notag
 G_{\mathbf{k}} = 2\,\gamma\,\left( \begin{array}{cccc}      
 \,\, 0 \quad\quad & 0 & 0 & 0 \\
 \,\, 0 \quad\quad & k_x^2-1 & k_x\,k_y & k_x\,k_z \\
 \,\, 0 \quad\quad & k_x\,k_y & k_y^2-1 & k_y\,k_z \\
 \,\, 0 \quad\quad & k_x\,k_z & k_y\,k_z & k_z^2-1 
 \end{array}\right) ,
\end{equation}
%\begin{equation}
%\widetilde{G}_0+\widetilde{G}_{\vec{k}}  =  
%\left(
%\begin{array}{cccc}
%0 & \,\, 0 & \,\,\, 0 & 0 \\ 
%\begin{array}{c} 0 \vphantom{G_{\vec{k}}} \\ 0  \vphantom{G_{\vec{k}}} \\ 0\vphantom{G_{\vec{k}}} \end{array}
%& \multicolumn{3}{c}{ \boxed{ \begin{array}{c} \\ G_0+G_{\vec{k}} \\ \, \end{array} } }
%\end{array}
%\right)
%\end{equation}
\begin{equation} \notag
G_1  =  -\,\frac{D}{2}
\left(
\begin{array}{cccc}
1 + \frac{\delta\Omega^2}{\Omega^2}  & 0 & 0 & 2\,\frac{\delta\Omega}{\Omega} \\
0 & 1 + \frac{\delta\Omega^2}{\Omega^2}  & 0 & 0 \\
0 & 0 & 1 + \frac{\delta\Omega^2}{\Omega^2} & 0\\
2\,\frac{\delta\Omega}{\Omega} & 0 & 0 & 1 + \frac{\delta\Omega^2}{\Omega^2}
\end{array}
\right) ,
\end{equation}
\begin{equation} \notag
G_2  =  \frac{D}{2}
\left(
\begin{array}{cccc}
1 + \frac{\delta\Omega^2}{\Omega^2}  & 0 & 0 & 2\,\frac{\delta\Omega}{\Omega} \\
0 & 1 - \frac{\delta\Omega^2}{\Omega^2}  & 0 & 0 \\
0 & 0 & 1 - \frac{\delta\Omega^2}{\Omega^2} & 0\\
2\,\frac{\delta\Omega}{\Omega} & 0 & 0 & 1 + \frac{\delta\Omega^2}{\Omega^2} 
\end{array}
\right) .
\end{equation}
The above equation is obtained to lowest order in $\rho \Omega^2 \ll V_d$,
under the assumptions that: (i) the detector's transition amplitude
only weakly depends on the energies, that is, $\Omega_{lr} \equiv \Omega$; (ii) 
the densities of states in the QPC's collectors are constant,
$\rho_l(E_{l_k}) \equiv \rho_l$ and $\rho_r(E_{r_k}) \equiv \rho_r$; (iii) 
 at $t=0$  the energy levels of the detector are filled up to the
Fermi-level so that  $m=0$, that is, $\mathbf{v}^{(n)}(0) = (1, v_x, v_y,
v_z) \cdot \delta_{n,0}$.  
As evident in Eq.~(\ref{eq:ODE}), for $D \rightarrow 0$ or
$\delta\Omega \rightarrow 0$, respectively, the system evolves
undisturbed, while for $\gamma \rightarrow 0$ our result reduces to that in Ref.~\onlinecite{Gurvitz:1997}.

According to Eqs.~(\ref{eq:weak_current}) and (\ref{eq:perturbation_omega})
we can obtain the 
expression for the WV by solving the differential
equations~(\ref{eq:ODE}) perturbatively in the regime $\delta\Omega \ll
\Omega$. 
The details of the
derivation are given in
Appendix~\ref{sec:appendix_perturbative_solution}. 
We highlight here that the perturbative solution is a valid
approximation for $\tau \ll 1/D (\Omega/\delta \Omega)^2$ exactly
corresponding to the weak measurement regime.
This finally yields the expression for the WV
\begin{equation}\label{eq:same_result}
 \left.\vphantom{\langle \Omega \rangle}\right._{\Pi_{\textrm{f}}}\langle I(\tau) \rangle^{\textrm{weak}}_{\Psi(t_0)} = e\,D \,  \frac{ \frac{1}{\tau} \, \intop_{0}^{\tau} \textrm{d}s \, v_z(s) + \frac{1}{\tau} \, \intop_{0}^{\tau} \textrm{d}s \, n_z(s)}{\mathbf{n}\cdot \mathbf{v}(\tau)} .
\end{equation}
Here $\tau$ is the duration time of the weak measurement and we
effectively introduced a time dependence in the  postselection
operator $\Pi_f(\tau) = 1/2 \, (\mathbf{n}(\tau) \cdot \mathbf{\sigma})$,
$\mathbf{n}=\mathbf{n}(\tau)$ is then the postselected state at time
$t=\tau$ instantaneously after the measurement. 
In the notation in Eq.~(\ref{eq:same_result}), $\mathbf{v}(s)$ is defined by $\mathbf{v}(s) = \exp\left[ \left( G_0 + G_{\mathbf{k}}  \right) s \right] \cdot \mathbf{v}(0)$, while $\mathbf{n}(s) = \exp\left[ \left(-G_0 + G_{\mathbf{k}}  \right)  s \right] \cdot \mathbf{n}(0)$.

The result in Eq.~(\ref{eq:same_result}) is already captured by a minimal model where the coupling to the detector is described by a von Neumann Hamiltonian~\cite{Neumann} $H_{\textrm{int}}(t) = \lambda g(t)\, \hat{p}\, \sigma_z$. It linearly couples the measured observable $\sigma_z$ to a detector's variable $\hat{p}$, $\lambda$ indicates the coupling constant and the time dependency of the interaction is included in the function $g(t)$. The WV of $\sigma_z$ is inferred from the conditional value of the conjugate variable $\hat{q}$,
\begin{equation}\label{eq:correlation}
 \left.\vphantom{\langle \Omega \rangle}\right._{\Pi_{\textrm{f}}}\langle \hat{q}(\tau) \rangle^{\textrm{weak}}_{\Psi(t_0)} = \lambda \, \textrm{Re}  \int_0^{\tau} \textrm{d}s \, \frac{\langle \Pi_f(\tau) \, \sigma_z(s) \rangle }{\langle \Pi_f(\tau) \rangle }   
\end{equation}
obtained to leading order in the coupling. The effect of decoherence is included in the correlation function $\langle \Pi_f(\tau) \, \sigma_z(s) \rangle$ resulting in Eq.~(\ref{eq:same_result}). Equation~(\ref{eq:correlation}) elucidates the role of \emph{coherent} system evolution between the measurement at time $s$ and the postselection at time $\tau$.

\subsection{Continuous readout}\label{sec:continuous_read-out}

Opposed to a coherent detection, the regime of continuous readout
is characterized by the detector's state being frequently read out by
the pointer.
This limit is described by a sequence of readouts at times $t=t_k$,
$k=1 \ldots N$, where the time interval between readouts $\Delta t_k
:= t_{k+1} - t_{k} \equiv \Delta t$ is the shortest time scale in the
problem, that is, $\Delta t \ll \textrm{min} \{1/\omega, 1/\gamma, 1 /D \}$. 
The conditional number of transmitted  electrons
Eq.~(\ref{eq:weak_current}) can now be expressed as the sum over all
permutations describing quantum jumps at all possible times
\begin{equation}\label{eq:permutations_1}
\left.\vphantom{\langle \Omega \rangle}\right._{\Pi_{\textrm{f}}}\langle n(\tau) \rangle_{\Psi(t_0)} =  \\ 
\frac{ \sum\limits_{\sum_j m_j = m} m \,\textrm{tr}\left\{ \Pi_f \cdot \langle \mathcal{R}^{(m_1,\ldots,m_N)}(\tau)\rangle_{\xi} \right\}  }{ \sum\limits_{\sum_j m_j = m} \textrm{tr}\left\{ \Pi_f \cdot \langle \mathcal{R}^{(m_1,\ldots,m_N)}(\tau)\rangle_{\xi} \right\}  } .
\end{equation}
%j=1,...,n
where $j=1 \ldots N$ and $\mathcal{R}^{(m_1,\ldots,m_N)}(\tau)$ characterizes the qubit's density matrix weighted with the probability that exactly $m_k$ electrons have passed within each time interval $\Delta t_k$. Each readout corresponds to a collapse of the qubit-detector system in the sector of $m_k$ electrons having passed with
the corresponding probability $P(m_k;t_k)$.

In the regime $\Delta t \ll \textrm{min} \{1/\omega, 1/\gamma, 1 /D \}$ at most one electron penetrates through the QPC between two subsequent readouts.~\footnote{While we are neglecting the terms with more than one electron having been transferred, the probability is still exactly conserved at any time since $tr\{\rho^{\textrm{(system)}}(t)\} = \sum_{n=0}^{\infty} v_0^{(n)}(t) \equiv 1$.} %foot2 (see appendix!).
The probabilities that either exactly one (quantum jump~\cite{Korotkov:2001,Korotkov_Averin:2001}) or zero electrons  accumulate in the collector within a readout period time are computed in appendix~\ref{sec:permutation}. They are given by $ P(0;\Delta t) = \{A \cdot \mathbf{v}^{(n_k)}(t_k)\}_0$ and $ P(1;\Delta t) = \{B \cdot \mathbf{v}^{(n_k)}(t_k)\}_0$, respectively, where the index $\{\ldots\}_0$ denotes the zeroth component and
\begin{align}\label{eq:def_A_and_B}
 A &= {1}_{(4)} + \left( G_0 + G_{\mathbf{k}} + G_1 \right)\,\Delta t + \mathcal{O}\left( \Delta t^2 \right )  \notag\\
 B &= G_2 \Delta t + \mathcal{O}\left( \Delta t^2 \right ). 
\end{align}
$ \mathcal{O}( \Delta t^2 )$ indicates higher order terms for $\Delta t \ll \textrm{min} \{1/\omega, 1/\gamma, 1 /D \}$. In the limit of $N \rightarrow \infty$, $\Delta t \rightarrow 0$ while keeping $N \cdot \Delta t = \tau$ constant, the conditional number of transferred electrons in Eq.~(\ref{eq:permutations_1}) can be analytically evaluated (cf. appendix~\ref{sec:permutation}), yielding the exact result
\begin{equation}\label{eq:n_multiple_readouts}
 \left.\vphantom{\langle \Omega \rangle}\right._{\Pi_{\textrm{f}}}\langle n(\tau) \rangle_{\Psi(t_0)} = \tau\, \frac{                     \mathbf{n}   \cdot   G_2   \cdot       e^{(G_0 + G_{\mathbf{k}}+ G_1 + G_2 ) \,\tau} \cdot \mathbf{v}(0)                  }{            \mathbf{n}   \cdot         e^{(G_0 + G_{\mathbf{k}}+ G_1 + G_2 ) \,\tau} \cdot \mathbf{v}(0)                }.
\end{equation}
The WV can be easily extracted by $_{\Pi_{\textrm{f}}}\langle I(\tau)
 \rangle^{\textrm{weak}}_{\Psi(t_0)} = \lim_{\delta\Omega/\Omega \to 0} (_{\Pi_{\textrm{f}}}\langle I(\tau)
 \rangle_{\Psi(t_0)}  )$, where
\begin{equation}\label{result-continuous}
 \left.\vphantom{\langle \Omega \rangle}\right._{\Pi_{\textrm{f}}}\langle I(\tau) \rangle_{\Psi(t_0)}^{\textrm{weak}} =  \frac{e\Omega}{\delta\Omega} \left(    \frac{1}{\tau} \cdot    \vphantom{\rangle}_{\Pi_{\textrm{f}}}\langle n(\tau) \rangle_{\Psi(t_0)} -   \frac{D}{2}   \right)   + \mathcal{O}\left( \frac{\delta\Omega^2}{\Omega^2} \right )                
\end{equation}
The simultaneous presence of $G_1 + G_2$ and $G_{\mathbf{k}}$ in
Eq.~(\ref{eq:n_multiple_readouts}) defines a new time scale
$1/\gamma \simeq 1/D (\Omega/\delta\Omega)$ which describes
the extra source of decoherence emerging from the detector itself. 
Note also that
Eqs.~(\ref{eq:n_multiple_readouts}) and (\ref{result-continuous}) are valid
at any time. 
The derivation indeed  relies on the perturbative solution in
Appendix~\ref{sec:appendix_perturbative_solution}, which is valid in
the limit $\Delta t \to 0$; hence, the exact composition of subsequent
evolution between different readouts holds at any time.

\section{Results}\label{sec:results}

Equations.~(\ref{eq:same_result}), (\ref{eq:n_multiple_readouts}) and (\ref{result-continuous}) represent the
main results of our paper. 
They express the WVs in the two limiting cases of coherent
detection and continuous readout in terms of the system dynamics.
Generally, they give rise to four different time scales, which characterize (i) the system's
dynamics, $1/\omega$, (ii) the decoherence, $1/\gamma$, (iii) the
detector dynamics, $1/D$, and (iv) the backaction, $1/D
\cdot (\Omega/\delta\Omega)^2$. We realistically
assume $1/D$ to be the shortest time scale in our problem and focus on $\tau \gg 1/D$. The effects of the detector at this time scale dominated by the Zeno effect,~\cite{Gurvitz:2008} though inherent in Eq.~(\ref{eq:ODE}), do not play a significant role at the larger time scales of interest where decoherence takes place. 
Consistently with our perturbative analysis, we further consider $ \tau \ll 1/D
\cdot (\Omega/\delta\Omega)^2$.

\begin{figure}[t]
		\hfill \subfigure[]{\includegraphics[width=0.18\textwidth]{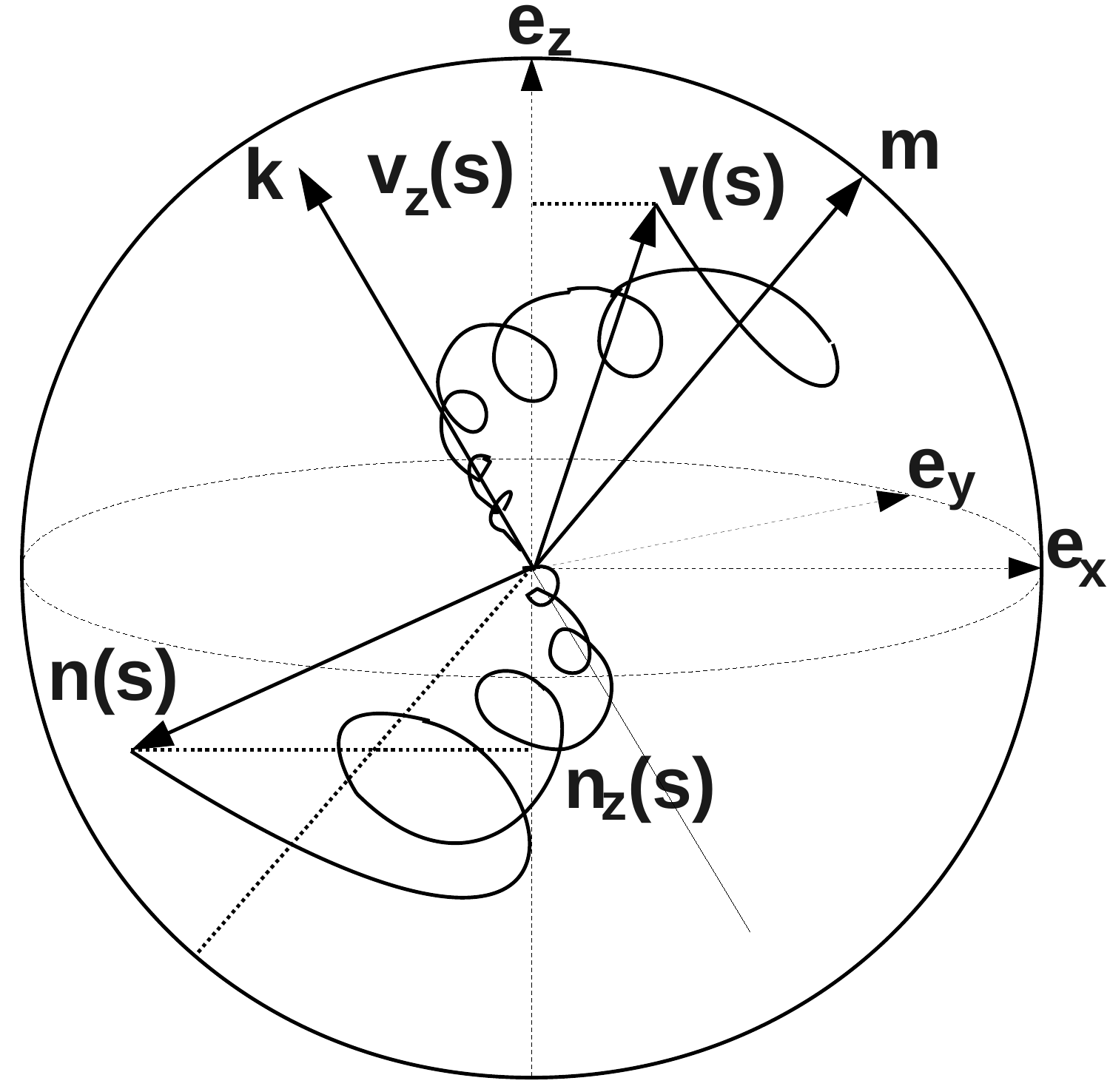}}\hfill
		\subfigure[]{\includegraphics[width=0.18\textwidth]{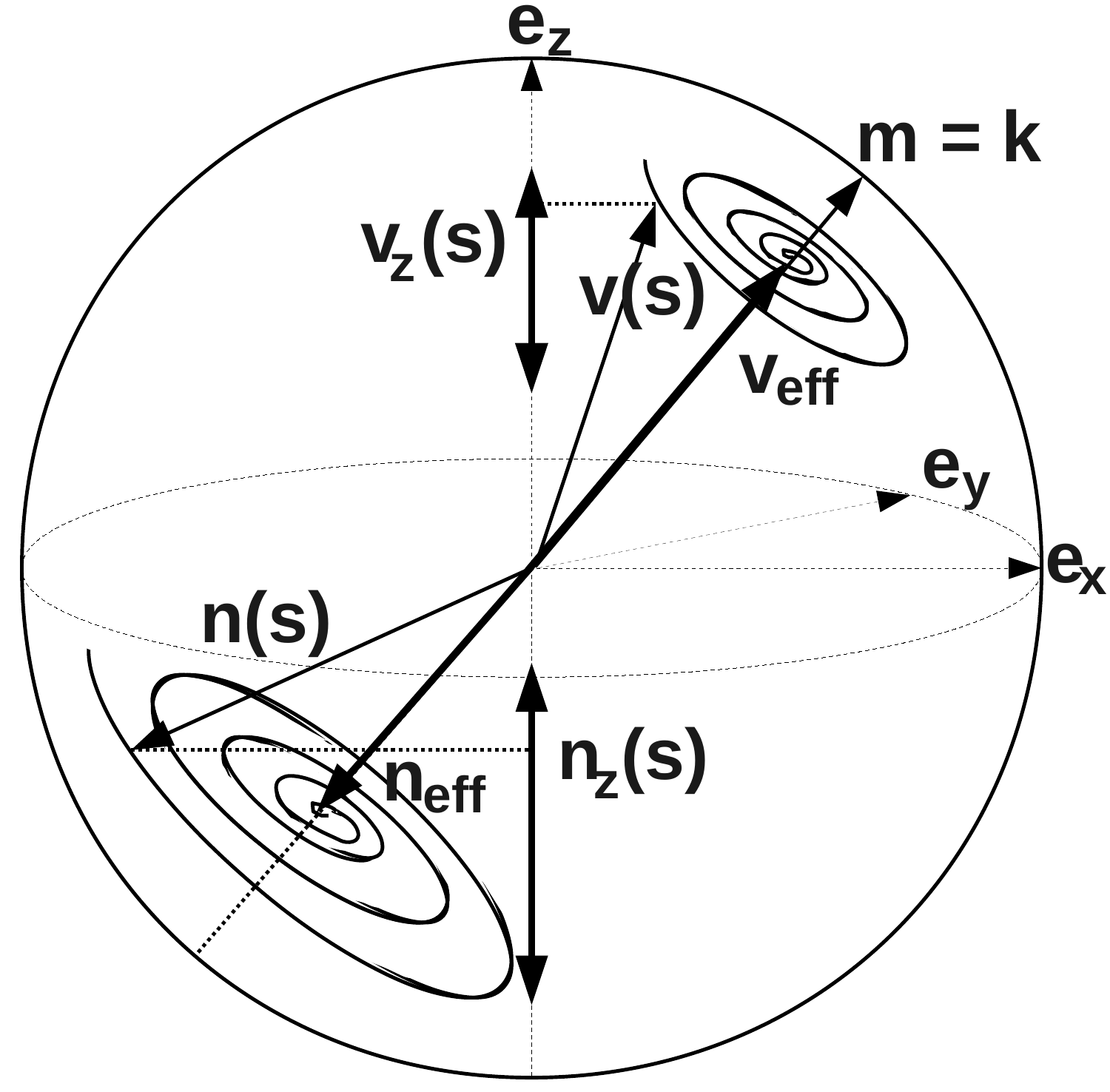}}\hfill
                \subfigure[]{\includegraphics[width=0.235\textwidth]{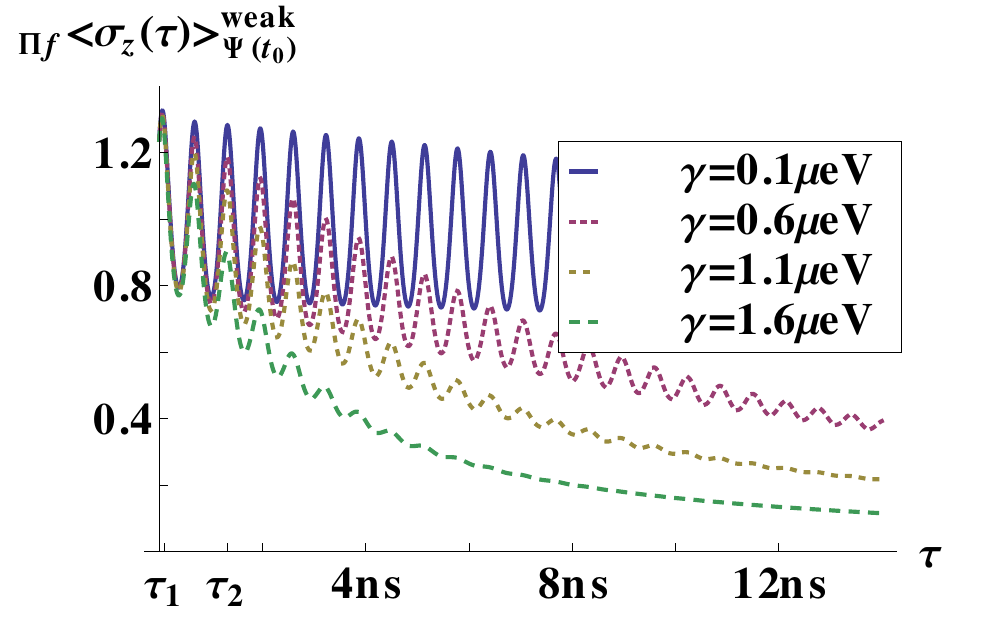}}
		\subfigure[]{\includegraphics[width=0.235\textwidth]{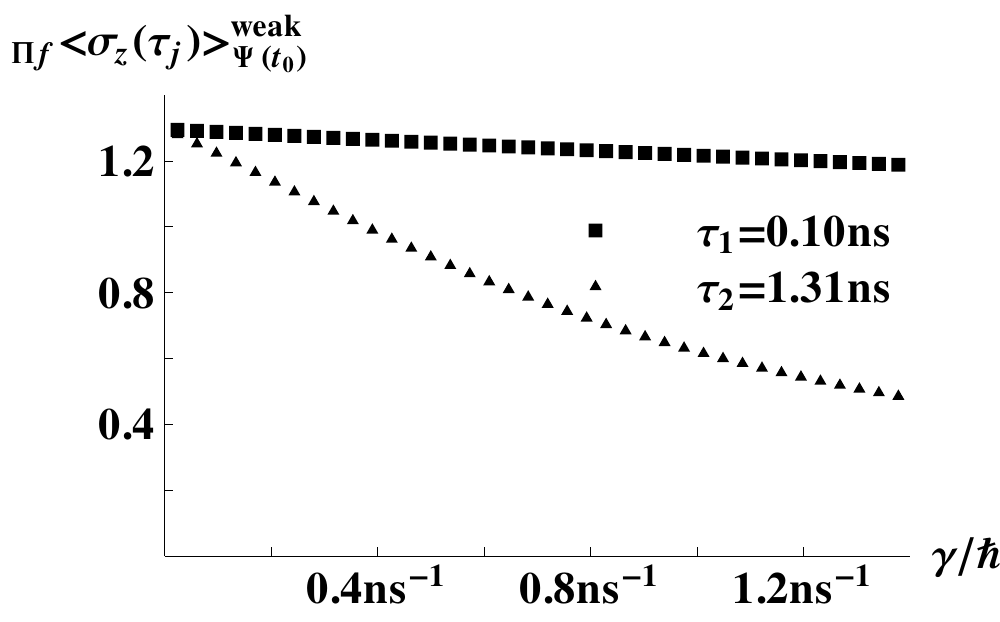}}		
		\caption{Schematic evolution of $\mathbf{v}(s)$ and
                  $\mathbf{n}(s)$ on the Bloch-sphere for  (a)
                  $\mathbf{m} \neq \mathbf{k}$ and (b) $\mathbf{m} =
                  \mathbf{k}$. (c) WV as a function of $\tau$
                  for different $\gamma$; (d) WV for different
                  $\gamma$ at $\tau_1 = 0.10\,\textrm{ns}$ and $\tau_2
                  = 1.31\,\textrm{ns}$,  marked in panel (c). In all
                  plots the parameters are chosen as $\epsilon_0 =
                  20\,\mu\textrm{eV}$, $\Delta_0 = 3\,\mu\textrm{eV}$, $\mathbf{k} \approx
                  -0.89\,\mathbf{e}_x + 0.45\, \mathbf{e}_z$, 
                  $\mathbf{v}(0)= (0.5, -0.33, 0.80)$ and $\mathbf{n}= (-0.20, 0.75, 0.63)$.} 
	\label{fig3}
\end{figure}

 The WV can be visualized via the motion of $\mathbf{v}(t)$ and
$\mathbf{n}(t)$ on the Bloch-sphere for $\mathbf{v}=(v_x, v_y, v_z)$ and
$\mathbf{n}=(n_x, n_y, n_z)$, as depicted in
Fig.~\ref{fig3}. $|\mathbf{v}(t)| \leq 1$ characterizes the
coherence of the qubit's state, which is initially prepared in a pure
quantum state with $|\mathbf{v}(t=0)|=1$.

We start with analyzing the case of coherent
detection, that is, the dynamical evolution obtained in
Eq.~(\ref{eq:same_result}). 
In this regime the effect of decoherence essentially reduces to the
dynamics of the qubit alone in the presence of decoherence. 
Accordingly, the WV presents different behaviors in different
regimes defined by the durations of the
measurement, $\tau$, compared to the remaining time scales
$1/\omega$, $1/\gamma$.

For long measurement durations $\tau \gg 1/\gamma$, the qubit's state generally relaxes towards a fully
statistical mixture, $\mathbf{v}(t)=0$, as shown in Fig.~\ref{fig3}(c), except for the special case
$\vec{k} \parallel \vec{m}$, which is discussed later.
Consequently, at time scales where decoherence effects come to
play we obtain $\textrm{max} | \left.\vphantom{\langle \Omega
         \rangle}\right._{\Pi_{\textrm{f}}}\langle I(\tau)
     \rangle^{\textrm{weak}}_{\Psi(t_0)} | \leq e D  $
and, hence, the peculiar characteristics of WVs are washed out.

For measurements shorter than the decoherence time, the results depend
on the system's evolution time scale.
For short measurements, $\tau \ll 1/\omega$,  which correspond to
negligible dynamics and fluctuations, the vectors $\mathbf{v}(s)$ and $
\mathbf{n}(s)$ in Eq.~(\ref{eq:same_result}) are constant so that a measurement of the averaged
detector's response trivially reflects the WV of the
observable $\hat{\sigma}_z$ independent from the measurement duration
time $\tau$, $\left.\vphantom{\langle \Omega
    \rangle}\right._{\Pi_{\textrm{f}}}\langle \sigma_z(\tau
\rightarrow 0) \rangle^{\textrm{weak}}_{\Psi(t_0)} = \textrm{Re}(
\,\left\langle\, \textrm{f} \,\right|  \hat{\sigma}_z \left|
    \Psi \right\rangle/\left\langle \,\textrm{f}\, \right|
  \left. \Psi \right\rangle\,)$.

The system's dynamics for intermediate durations of the measurement, $\tau \gg
1/\omega$, however, are appreciable. Here, both
vectors $\mathbf{n}(s)$ and $\mathbf{v}(s)$  precess about the
eigenvector $\mathbf{m}$ of $G_0$ with a frequency of $2\, \omega$.
Note that, due to its backward-in-time evolution,
$\mathbf{n}(s)$ precesses in the opposite direction as compared to
$\mathbf{v}(s)$.
Due to the oscillatory dependence of the denominator in Eq.~(\ref{eq:same_result}),
peculiar WVs may occur for properly fine-tuned measurement
duration times. WVs much larger than $1$ are realized, for instance, for orthogonal states when $|1+\mathbf{n} \cdot \mathbf{v}(\tau)| \gg 1$, which leads to  $\tau = n \cdot \frac{\pi}{\omega} \pm \Delta \tau$, with $\Delta\tau \ll \frac{1}{|\sin^2 \alpha|} \, \frac{1}{2\,\omega}$,
%\begin{equation}
% \tau = n \cdot \frac{\pi\,\hbar}{\omega} \pm \Delta \tau \quad \textrm{with}\quad \Delta\tau \ll \frac{1}{|\sin^2 \alpha|} \, \frac{\hbar}{2\,\omega}
%\end{equation}
where $\alpha$ denotes the altitude angle between $\mathbf{v}(0)$ and
$\mathbf{m}$. The effect of the strength of the fluctuations on short and intermediate measurements is
shown in Figs.~\ref{fig3}(c) and~\ref{fig3}(d), resulting in a trivial decay towards the
steady-state value.
\begin{figure}[t]
	\centering
		\subfigure[]{\includegraphics[width=0.235\textwidth]{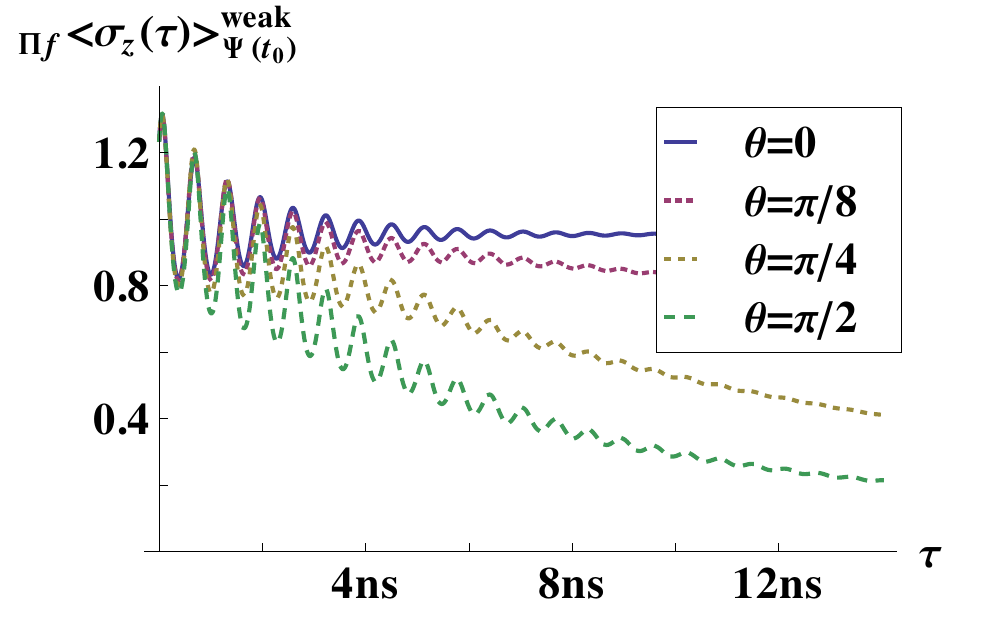}}
		\subfigure[]{\includegraphics[width=0.235\textwidth]{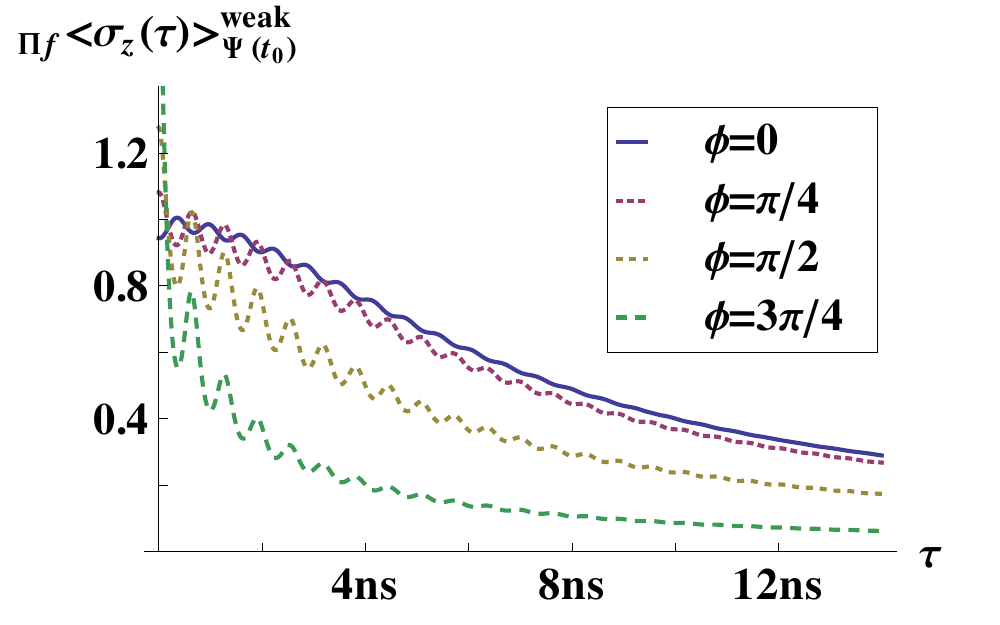}}
		\subfigure[]{\includegraphics[width=0.235\textwidth]{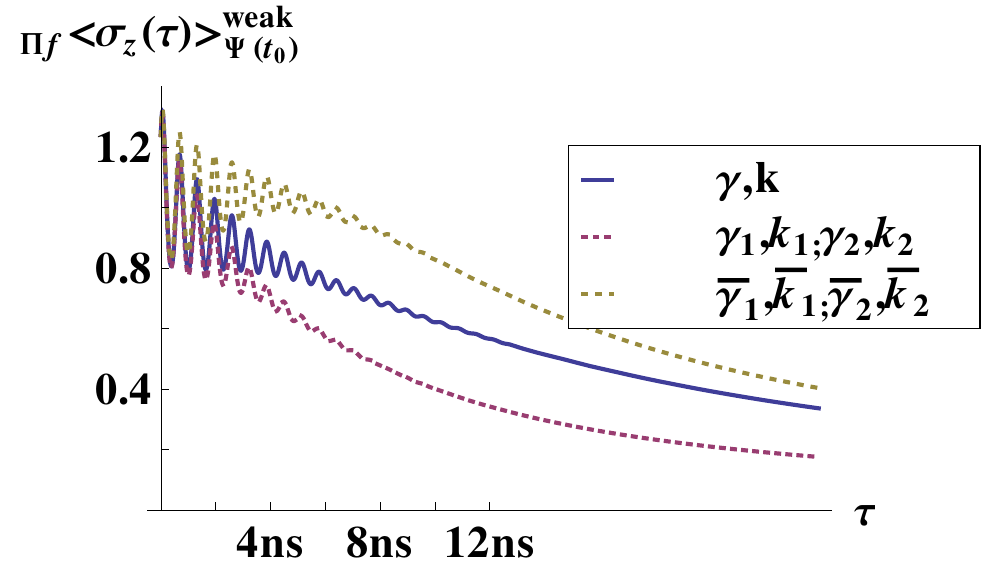}}
		\caption{WV as a function of $\tau$ for (a) different $\mathbf{k}$
($\theta = 0$ corresponds to $\mathbf{k} \parallel \mathbf{m}$) and
(b) different $\mathbf{n}$ ($\phi = 0$ corresponds to
$\mathbf{n} \parallel \mathbf{m}$). (c) Same as in (a) in the presence of multiple decoherence sources, where
                  $\gamma=1.0\,\mu\textrm{eV}$, $\mathbf{k} =
                  \frac{1}{\sqrt{2}}\,\mathbf{e}_x +
                  \frac{1}{\sqrt{2}}\, \mathbf{e}_z$, 
                  $\mathbf{k}_1 = \frac{1}{\sqrt{2}}\,\mathbf{e}_x$,
                  $\mathbf{k}_2 = \frac{1}{\sqrt{2}}\, \mathbf{e}_z$,
                  $\gamma_1= \frac{1}{\sqrt{2}}\,\mu\textrm{eV}$
                  $\gamma_2=\frac{1}{\sqrt{2}}\,\mu\textrm{eV}$, 
                  $\overline{\mathbf{k}}_1 = \mathbf{m}$,
                  $\overline{\mathbf{k}}_2 = \mathbf{m}_{\perp}$,
                  $\overline{\gamma}_1= (\mathbf{k} \cdot
                  \mathbf{m})\,\mu\textrm{eV}$, 
                  $\overline{\gamma}_2=(\mathbf{k} \cdot
                  \mathbf{m}_{\perp})\,\mu\textrm{eV}$; all the other parameters are as in Fig.~\ref{fig3}.}
	\label{fig4}
\end{figure}

As already anticipated, an important role is played by the
``direction'' of the noise term. Its effect is illustrated in
Fig.~\ref{fig4}(a). As long as $\mathbf{k}$ becomes more and more
parallel to the $\mathbf{m}$, the relaxation time toward the steady
state becomes longer.
 In the limiting case, it relaxes to a partially coherent state.
The direction of the postselection orientation plays a similar role,
as shown in Fig.~\ref{fig4}(b). 
We also note that, in the presence of multiple decoherence sources, even for
 a given effective direction and strength of the fluctuating term, the
decay of the WV still strongly depends on the relative
directions between different sources [cf. Fig.~\ref{fig4}(c)].

 \begin{figure}[t]
		\subfigure[]{\includegraphics[width=0.235\textwidth]{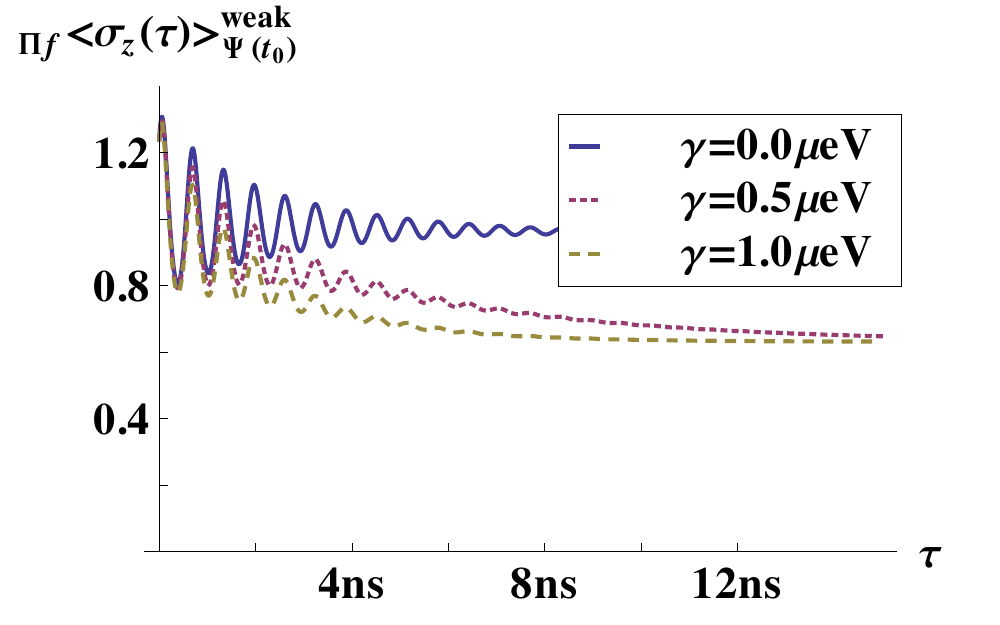}}
		\subfigure[]{\includegraphics[width=0.235\textwidth]{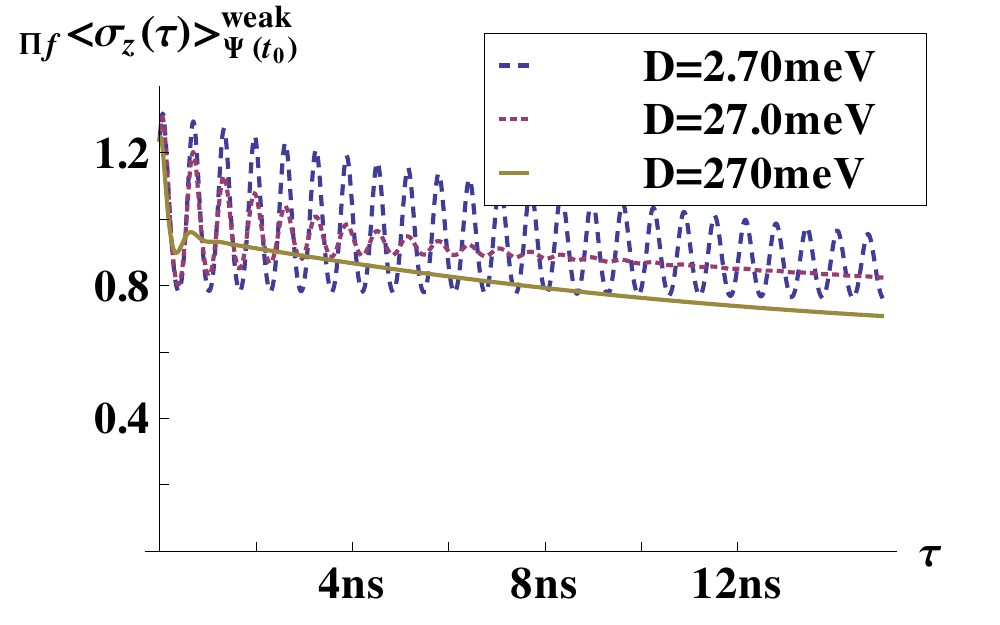}}
		\caption{WV for continuous detection for (a) different $\gamma$ with $D=0.270\,\textrm{eV}$ and (b) different $D$ with $\gamma = 0.1\,\mu\textrm{eV}$, other parameters as in Fig.~\ref{fig3}; in all plots $\delta\Omega/\Omega = 0.001$.}
	\label{fig5-6}
\end{figure}

\begin{figure}[t]
		\subfigure[]{\includegraphics[width=0.235\textwidth]{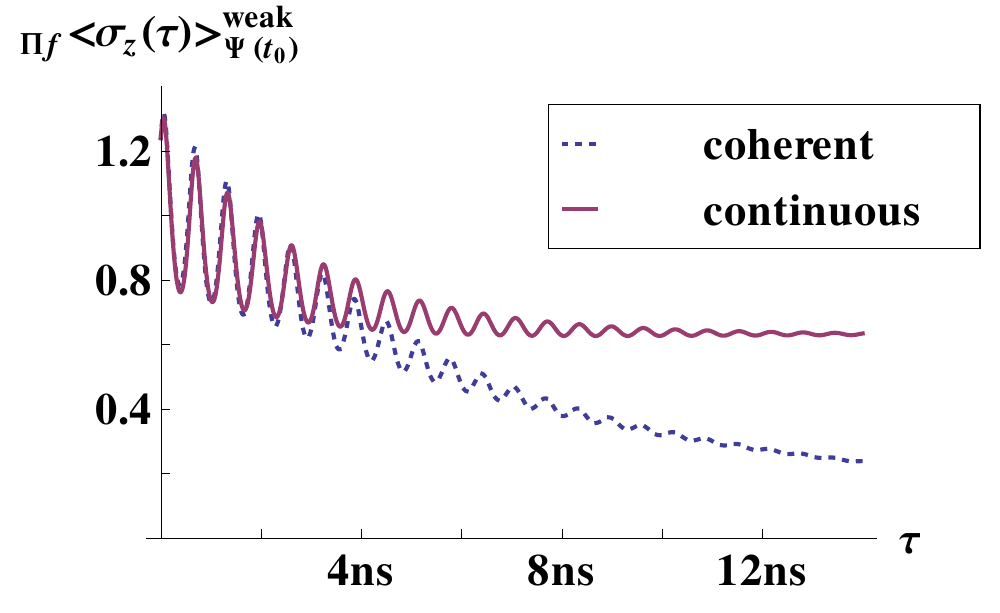}}
		\subfigure[]{\includegraphics[width=0.235\textwidth]{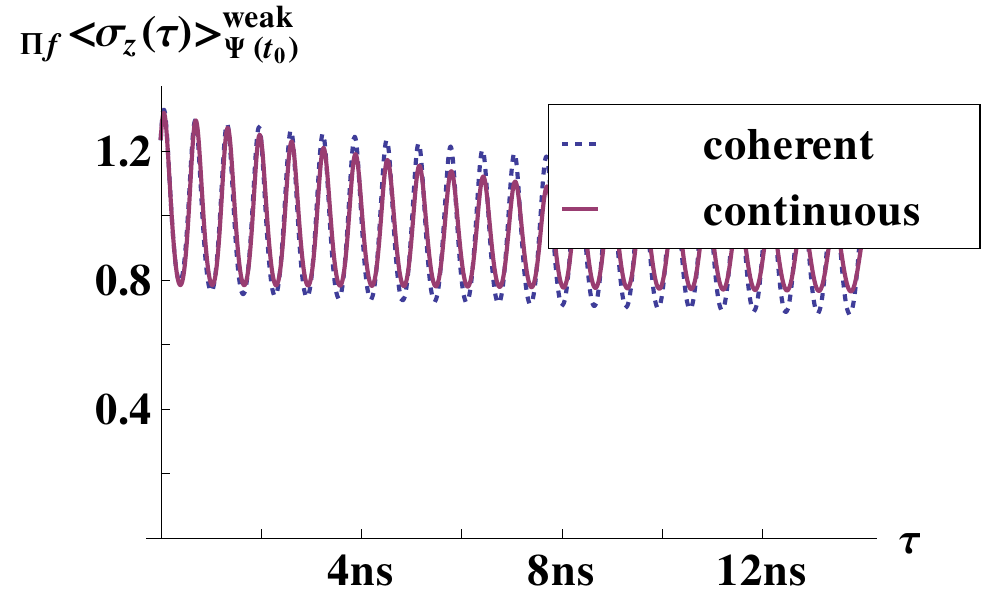}}
		\subfigure[]{\includegraphics[width=0.235\textwidth]{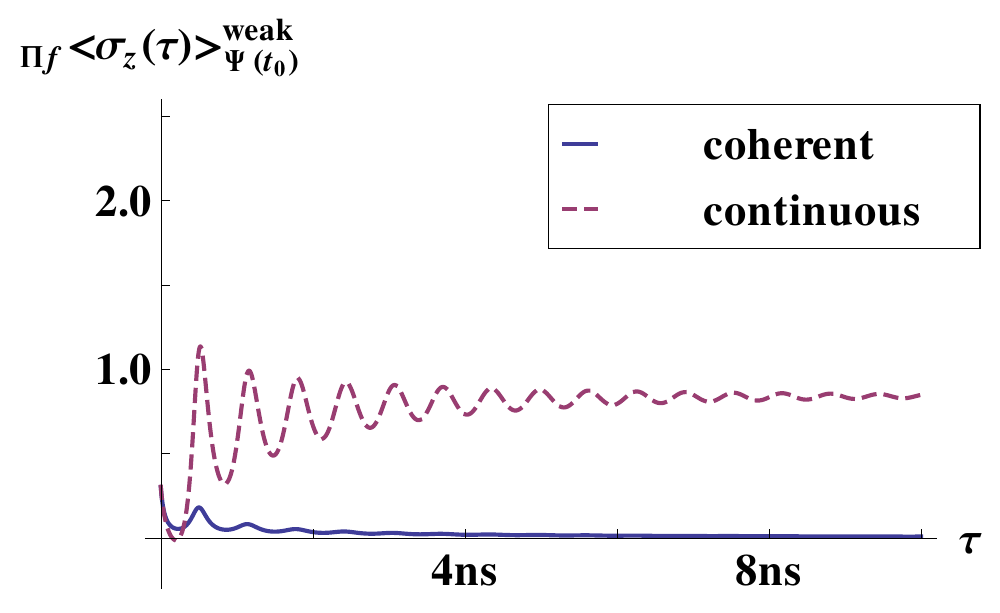}}
		\subfigure[]{\includegraphics[width=0.235\textwidth]{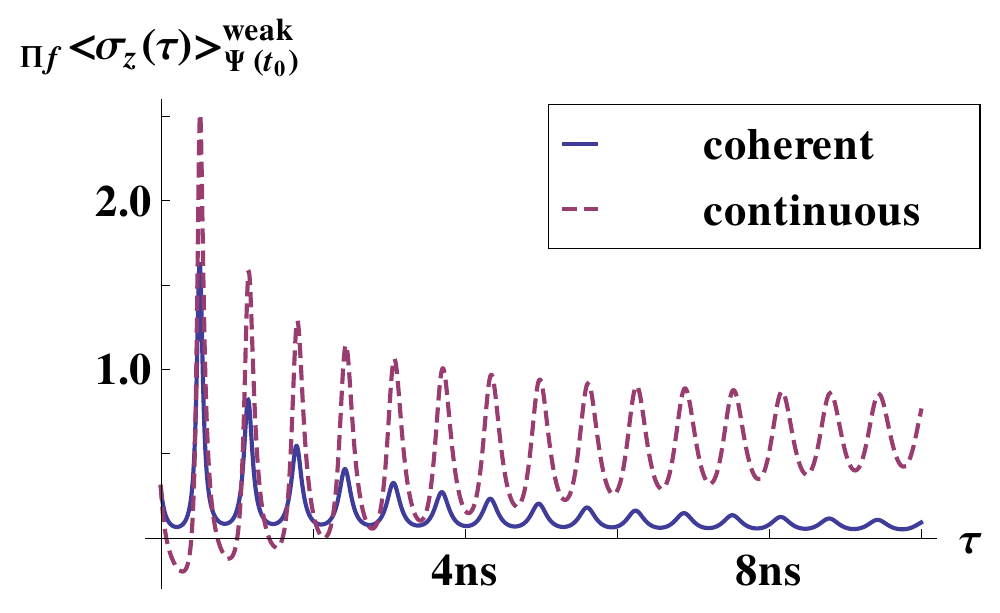}}
		\caption{Weak value for continuous detection and coherent readout for different $\gamma$ and pre- and postselection states $\mathbf{n}$, $\mathbf{v}$; $\gamma = 1.0\,\mu\textrm{eV}$ in (a), (c) and $\gamma = 0.1\,\mu\textrm{eV}$ in (b), (d); $\mathbf{v}(0)= (0.5, -0.33, 0.80)$, $\mathbf{n}= (-0.20, 0.75, 0.63)$ in (a,b) and $\mathbf{v}(0)= (-0.5, 0.33, -0.80)$, $\mathbf{n}= (0.50, 0.20, 0.84)$ in (c), (d). In all plots $D=0.270\,\textrm{eV}$, $\delta \Omega/\Omega = 0.001$ and the other parameters as in Fig.~\ref{fig3}.}
	\label{fig5}
\end{figure}

The case of a continuous readout is characterized by a rather different behavior. While in both, the coherent and continuous readout regimes, a significant strong decoherence destroys the peculiarities of WVs [cf.~Fig.~\ref{fig5-6}(a)], in the latter the detector itself introduces decoherence on the time scale of its backaction. This is visualized in Fig.~\ref{fig5-6}(b). It also shows that $D$ effects the results at the time scales $\tau \ll 1/D (\Omega/\delta\Omega)^2$ relevant for WVs, in contrast to the coherent case.
The difference between the two cases is highlighted in Fig.~\ref{fig5}, comparing the two regimes for different decoherence strength and postselection. The WV in the continuous case can be enhanced as compared to the coherent measurement, leading to peculiar WVs, where a coherent measurement would not [cf. Figs.~\ref{fig5}(c) and (d)]. This effect depends on the chosen postselection [cf. Fig.~\ref{fig5}(a) vs. Fig.~\ref{fig5}(c); Fig.~\ref{fig5}(b) vs. Fig.~\ref{fig5}(d)] and is suppressed by decoherence [cf. Fig.~\ref{fig5}(c) vs Fig.~\ref{fig5}(d)].
We can understand these results by analyzing the asymptotic behavior after decoherence has taken place. Equation~(\ref{result-continuous})
gives a WV 
\begin{equation}
  \left.\vphantom{\langle \Omega \rangle}\right._{\Pi_{\textrm{f}}}\langle \sigma_z \rangle^{\textrm{weak}}_{\Psi(t_0)} = \frac{\textrm{tr} \{ (1 + \mathbf{n} \cdot \mathbf{\sigma})  \, \sigma_z(s) \, \rho \} }{ \textrm{tr} \{ (1 + \mathbf{n} \cdot \mathbf{\sigma})  \,  \rho  \} } = n_z   
\end{equation}
for a fully incoherent state $\rho \propto 1$. This is the WV of an incoherent state.~\cite{Romito-Gefen-Blanter:2008} The results from Eq.~(\ref{eq:same_result}) for the coherent case lead instead to $_{\Pi_{\textrm{f}}}\langle \sigma_z \rangle^{\textrm{weak}}_{\Psi(t_0)} = 0$ for $\rho \propto 1$. This difference arises because of the \emph{coherent} evolution in the correlation between measurement and postselection [cf. Eq.~(\ref{eq:correlation})] which does not take place in the continuous readout due to the frequent ``pointer`` readout. The different steady states are shown in Fig.~\ref{fig5}. Though the steady states correspond to times beyond the weak measurement regimes, their difference reflects at shorter time scales as well. There it leads to enhanced WVs exceeding the strong boundary in one case and not in the other (cf. Fig.~\ref{fig5}(c) and~\ref{fig5}(d)). This explains the sensitivity  of the effect to postselection (that counts the steady state of the continuous case) and to decoherence (that suppresses faster peculiar WVs within the standard range in both cases).

\begin{figure}[h]
	\centering
		\subfigure[]{\includegraphics[width=0.235\textwidth]{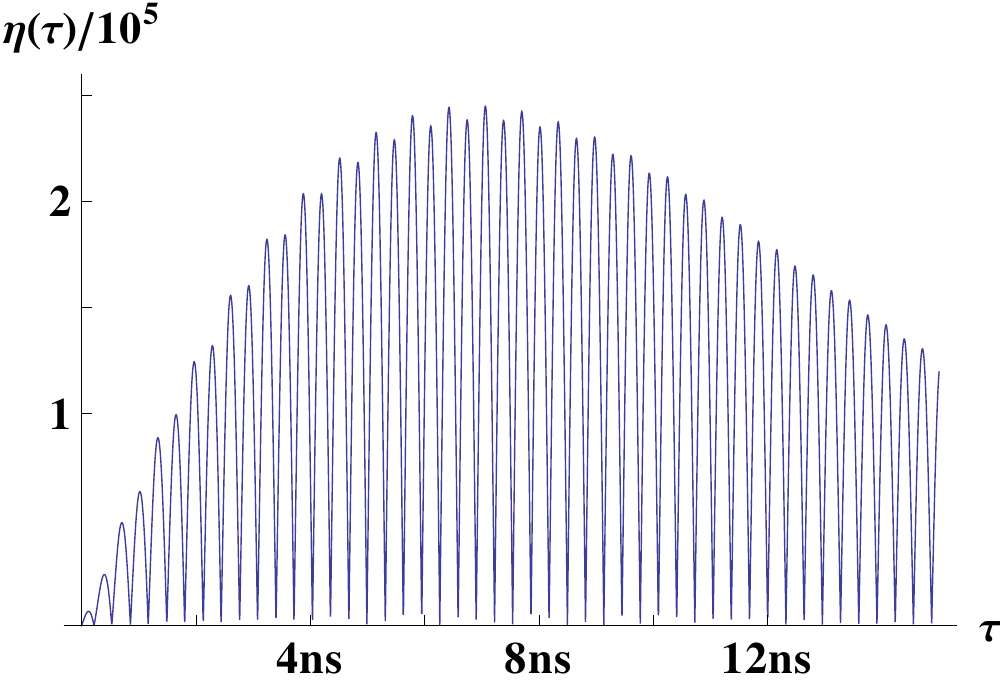}}
		\subfigure[]{\includegraphics[width=0.235\textwidth]{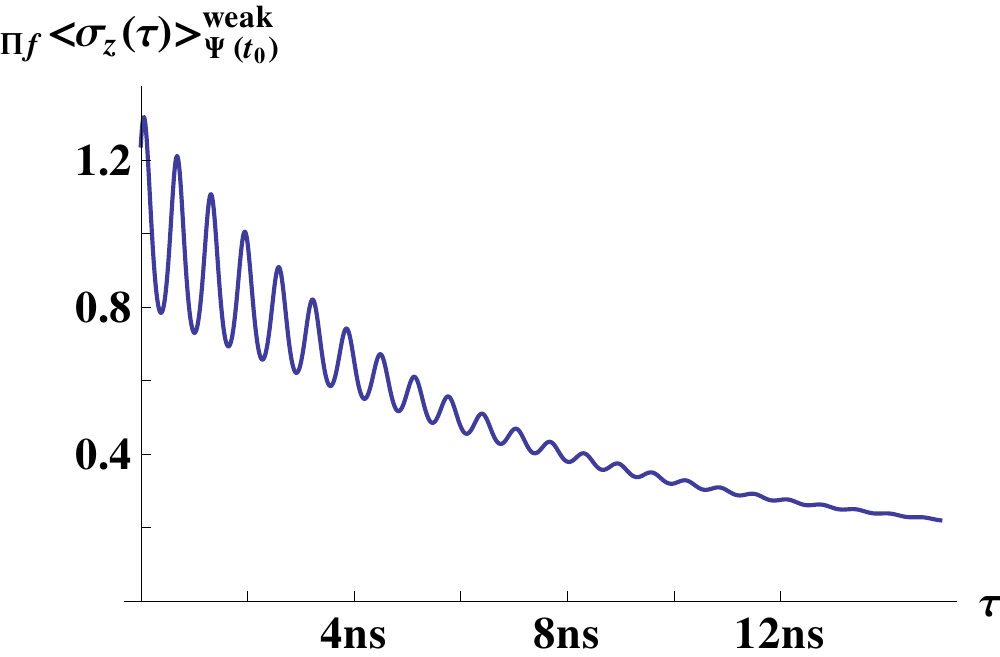}}
		\caption{Validity perturbation. (a) Left-hand side of Eq.~(\ref{eq:weakness_condition}) and (b) the corresponding WV for $\gamma = 1.0\,\mu\textrm{eV}$, $D=0.270\,\textrm{eV}$, $\delta \Omega/\Omega = 0.001$; all other parameters are as in Fig.~\ref{fig3}.}
	\label{fig6}
\end{figure}

The perturbative solution of the system's dynamics underlying the
result of the coherent detection allows us to discuss, to some extent,
the validity of the WV expression in Eq.(~\ref{eq:same_result}).
As a first check we can require that the second-order contribution is
irrelevant compared to the first-order contribution discussed so far,
that is,  $  | \left.\vphantom{\langle \Omega
    \rangle}\right._{\Pi_{\textrm{f}}}\langle I(\tau)
\rangle^{\textrm{weak(2)}}_{\Psi(t_0)} |  \cdot \delta\Omega^2 /
\Omega^2    \ll     | \left.\vphantom{\langle \Omega
    \rangle}\right._{\Pi_{\textrm{f}}}\langle I(\tau)
\rangle^{\textrm{weak}}_{\Psi(t_0)}  | \cdot \delta\Omega/ \Omega  $,
leading to the condition 
\begin{equation}\label{eq:weakness_condition}
\eta(\tau) =\frac{\delta \Omega}{\Omega}\cdot \frac{D\tau}{2}  \cdot
 \left| \frac{   v_x(\tau) \, n_x + v_y(\tau) \,
    n_y        }{ \frac{1}{\tau} \, \intop_{0}^{\tau}
    \textrm{d}s \, v_z(s) +  \frac{1}{\tau} \, \intop_{0}^{\tau}
    \textrm{d}s \, n_z(s)    } \right| \ll 1 \, .
\end{equation}
This imposes a restriction on the validity of the WV's result
also within the regime of weak measurement, as shown in Fig.~\ref{fig6}.
Indeed, for specific qubit's
 parameters $\epsilon_0$ and $\Delta_0$ and particular boundary
 conditions $v_z(0)$ and $n_z$, the numerator in
 Eq.~\eqref{eq:weakness_condition} vanishes
at finite duration times for $\tau \ll 1 / \gamma$ so that the
perturbation is valid at discrete times which depend on the chosen
 parameters (cf. Fig.~\ref{fig6}). 
For $\tau \gg 1 / \gamma $, on the contrary, $\eta \propto \tau^2
\,e^{-c\, \tau}$ where $c>0$ and the perturbation is
asymptotically valid %(cf. the inset in  Fig.~\ref{fig6}).
 unless $\mathbf{k} = \mathbf{m} = \mathbf{e}_x$.

\section{Conclusions}\label{sec:conclusion}
In this work we have addressed the effects of decoherence on weak
value measurements involving postselection. 
We have considered the paradigm model of a charge measurement in a
double quantum dot by a nearby QPC where we have
included fluctuations of the parameters due to external noise sources. 
After deriving a general expression for the postselected signal
(current) in the QPC in terms of the reduced density matrix of the
qubit, we have evaluated it explicitly in two different regimes
determined by the detector's readout, namely continuous vs ''single-time``
detector's readout. 

We have characterized the WV's dependence on the various
parameters of the system.
In particular, we have shown that statistical fluctuations of the qubit's parameters
generally reduce the WV into the classical range for
measurements longer than the decoherence time.
On shorter time scales we have determined a boundary for the region of
validity of the WV result.
Remarkably, there the continuous readout can lead to an enhancement of peculiar weak
values as compared to the case of coherent detection.

%\section*{Acknowledgments}
\acknowledgements
We would like to thank O. Ziberberg and Y. Gefen for very useful
discussions.
We acknowledge the financial support of ISF and the Minerva Foundation.

\begin{appendices}
 \numberwithin{equation}{section} %for equation numbering in the appendix as A.1, A.2 etc.

\section{Derivation of Averaged Rate Equations}\label{app:averaged_ODE}
%\numberwithin{equation}{section}

In this appendix we derive the differential equation~(\ref{eq:ODE}) out of Eqs.~(\ref{eq:ODEforv_z}) of the main text. In the following, the derivation is presented exemplarily only for $\hat{v}_x^{(m)} = \textrm{tr}\{\dot{\hat{v}}_{x}^{(m,m)}(t)\}$ since the other terms ($\hat{v}_y^{(m)}$, $\hat{v}_z^{(m)}$) are treated completely analogously. It is useful to perfom a ''Laplace'' transform for the whole matrices,
\begin{equation}
\hat{v}_{j}^{(m,n)}(E) = \lim_{\delta \to 0} \int\limits_{0}^{\infty} \hat{v}_{j}^{(m,n)}(t) \,\exp\left[ i\,(E + i \,\delta)\,t \right]\,\textrm{d}t\,,
\end{equation}
in order to include the initial conditions of the differential equations that $m=0$ electrons have penetrated through the collector at $t=0$ and, hence, the qubit-detector system is in a pure state. Here, $\delta > 0$ ensures the convergence of the integral. A high-energy cutoff is introduced concerning the inverse transform, so that the upper limit of the integral in the inverse transform is $\Lambda \rightarrow \infty$

We can write the differential equations for the Laplace transformed components, $\dot{\hat{v}}_{j}^{(m,m)}(t)$. In this regard, the matrix products in Eq.~(\ref{eq:ODEforv_z}) can be easily calculated by evaluating each $(m,n)$ block as introduced in Eq.~(\ref{eq:full_rho}). Due to the fact that the definition of  $\Delta \hat{H}^{+}$ includes terms proportional to $a_r^{\dagger}\,a_r$ and $a_l^{\dagger}\,a_l$, diagonal blocks of $\Delta \hat{H}^{+}$ are nonzero, whereas the diagonal blocks $\Delta \hat{H}^{- \, (m,m)}$ vanish. Moreover, since $\Delta \hat{H}^{\pm}$ consists of combinations $a_l^{\dagger}\,a_r + \textrm{h.c.}$ which raise or lower, respectively, the number of electrons in the detector by exactly one electron, only the off-diagonal blocks $\Delta \hat{H}^{\pm \, (m,m+1)}$ and $\Delta \hat{H}^{\pm \, (m+1,m)}$ neighboring the diagonal blocks are nonzero. Consequently, a blockwise evaluation of the matrix products can be written as
\begin{widetext}
\begin{align}
\left( \hat{v}_{j} \cdot \Delta \hat{H}^{-} \right)^{(m,n)} &= \hat{v}_{j}^{(m,n-1)} \cdot \Delta \hat{H}^{- \, (n-1,n)} + \hat{v}_{j}^{(m,n+1)} \cdot \Delta \hat{H}^{- \, (n+1,n)} \,, \\
\left( \hat{v}_{j} \cdot \Delta \hat{H}^{+} \right)^{(m,n)} &= \hat{v}_{j}^{(m,n-1)} \cdot \Delta \hat{H}^{+ \, (n-1,n)} +  \hat{v}_{j}^{(m,n)} \cdot \Delta \hat{H}^{+ \, (n,n)}  + \hat{v}_{j}^{(m,n+1)} \cdot \Delta \hat{H}^{+ \, (n+1,n)}\,,
\end{align}
 and analogous for $( \Delta \hat{H}^{+/-} \cdot \hat{v}_{j} )^{(m,n)}$. Note that $\hat{v}_{j}^{(m,k)} \cdot \Delta \hat{H}^{- \, (k,n)}$ still describes a product of matrices which are of infinite dimension. Thus, applying the "Laplace" transform to Eqs.~(\ref{eq:ODEforv_z}) for $j=x$ by considering exemplarily $\mathbf{k} = \mathbf{e}_z$  leads to
\begin{align}
\textrm{tr}& \left\{\dot{\hat{v}}_{x}^{(m,m)}(t) \right\} = - \pounds^{-1} \left[ \textrm{tr} \left\{ (i \, E - \delta)\,\hat{v}_{x}^{(m,m)}(E)  +  \hat{v}_{x}^{(m,m)}(0)  \right\}  \right]  \vphantom{\left[ \textrm{tr} \left\{    \underbrace{-2\,\gamma\,\hat{v}_{x}^{(m,m)}(E)}_{{A}}\right] \right\}}\notag\\
& =  \pounds^{-1} \left[ \textrm{tr} \left\{    \underbrace{-2\,\gamma\,\hat{v}_{x}^{(m,m)}(E)}_{=\hat{v}_a(E)} \underbrace{-\,2\,\epsilon_0\,\hat{v}_{y}^{(m,m)}(E)}_{=\hat{v}_b(E)}\right.\right. \notag\\
& \quad  -\,\frac{i}{2}\, \left( \underbrace{\Delta \hat{H}^{+ \, (m,m-1)} \cdot \hat{v}_{x}^{(m-1,m)}(E)}_{=\hat{v}_c(E)} + \underbrace{\Delta \hat{H}^{+ \, (m,m)} \cdot \hat{v}_{x}^{(m,m)}(E)}_{=\hat{v}_d(E)} + \underbrace{\Delta \hat{H}^{+ \, (m,m+1)} \cdot \hat{v}_{x}^{(m+1,m)}(E)}_{=\hat{v}_e(E)}  \right)                \notag\\
& \quad  +\,\frac{i}{2}\, \left( \underbrace{\hat{v}_{x}^{(m,m-1)}(E) \cdot \Delta \hat{H}^{+ \, (m-1,m)}}_{=\hat{v}_f(E)} +  \underbrace{\hat{v}_{x}^{(m,m)}(E) \cdot \Delta \hat{H}^{+ \, (m,m)}}_{=\hat{v}_g(E)} +  \underbrace{\hat{v}_{x}^{(m,m+1)}(E) \cdot \Delta \hat{H}^{+ \, (m+1,m)}}_{=\hat{v}_h(E)}  \right)          \notag\\
& \quad  -\,\frac{1}{2}\, \left( \underbrace{\Delta \hat{H}^{- \, (m,m-1)} \cdot \hat{v}_{y}^{(m-1,m)}(E)}_{=\hat{v}_i(E)} + \underbrace{\Delta \hat{H}^{- \, (m,m+1)} \cdot \hat{v}_{y}^{(m+1,m)}(E)}_{=\hat{v}_j(E)}   \right.  \notag\\
& \quad \quad + \left. \left. \left. \underbrace{\hat{v}_{y}^{(m,m-1)}(E) \cdot \Delta \hat{H}^{- \, (m-1,m)}}_{=\hat{v}_k(E)} + \underbrace{\hat{v}_{y}^{(m,m+1)}(E) \cdot \Delta \hat{H}^{- \, (m+1,m)}}_{=\hat{v}_l(E)}   \right)                 \right\} \right] \,,\label{eq:letters}
\end{align}
where $\pounds^{-1}$ denotes the inverse ``Laplace`` transform. We analyze the various terms separately. We start with observing that $\pounds^{-1}[ \hat{v}_a(E) ] = -2\,\gamma\,\textrm{tr} \{\hat{v}_{x}^{(m,m)}(t)\}$ and $ \pounds^{-1}[ \hat{v}_b(E) ] = -2\,\epsilon_0\,\textrm{tr} \{\hat{v}_{y}^{(m,m)}(t)\}$. Moreover, due to the cyclic invariance of the trace $\pounds^{-1}[ \hat{v}_d(E) ] +  \pounds^{-1}[ \hat{v}_g(E) ] = 0$. Employing the explicit expression of $\Delta \hat{H}^{\pm}$, reduces the evaluation of the remaining terms to the calculation of
 \begin{align}
\left( \Delta \hat{H}^{+ \, (m,m+1)}\right. \left. \cdot \hat{v}_{j}^{(m+1,n)} \right)_{(l_1 r_1 \ldots l_m r_m);(l'_1 r'_1 \ldots l'_n r'_n)} = \sum_{\substack{ l_{m+1} r_{m+1} }} (-1)^m 2\,\Omega_{l_{m+1} r_{m+1}}   \cdot \left( \hat{v}_{j}^{(m+1,m)} \right)_{(l_1 r_1 \ldots l_m r_m l_{m+1} r_{m+1});(l'_1 r'_1 \ldots l'_n r'_n)} 
\end{align}
and
 \begin{align}
\left( \Delta \hat{H}^{+ \, (m,m-1)}\right. \left. \cdot \hat{v}_{j}^{(m-1,n)} \right)_{(l_1 r_1 \ldots l_m r_m);(l'_1 r'_1 \ldots l'_n r'_n)} &= -\,2\,\Omega_{l_1 r_1} \cdot \left( \hat{v}_{j}^{(m-1,m)} \right)_{(l_2 r_2 \ldots l_m r_m);(l'_1 r'_1 \ldots l'_n r'_n)} \notag\\
& \quad+\,2\,\Omega_{l_2 r_2} \cdot \left( \hat{v}_{j}^{(m-1,m)} \right)_{(l_1 r_1 l_3 r_3 \ldots l_m r_m);(l'_1 r'_1 \ldots l'_n r'_n)}\notag\\
& \quad  -\,2\,\Omega_{l_3 r_3} \cdot \left( \hat{v}_{j}^{(m-1,m)} \right)_{(l_1 r_1 l_2 r_2 l_4 r_4 \ldots l_m r_m );(l'_1 r'_1 \ldots l'_n r'_n)}\notag\\
& \quad  \, \ldots \notag\\
& \quad + (-1)^m\,2\,\Omega_{l_m r_m} \cdot \left( \hat{v}_{j}^{(m-1,m)} \right)_{(l_1 r_1 \ldots l_{m-1} r_{m-1});(l'_1 r'_1 \ldots l'_n r'_n)}%\\
%& \hphantom{= \sum_{\substack{ l_{m+1} r_{m+1} }} (-1)^m \, 2\,\Omega_{l_{m+1} r_{m+1}} \cdot \left( v_{j,\textrm{QPC}}^{(m+1,m)} \right)_{(l_1 r_1 \ldots l_m r_m l_{m+1} r_{m+1});(l'_1 r'_1 \ldots l'_n r'_n)}}
\end{align}
for products concerning the off-diagonal boxes and
\begin{align}
\left( \Delta \hat{H}^{+ \, (m,m)}\right. & \left. \cdot \hat{v}_{j}^{(m,n)} \right)_{(l_1 r_1 \ldots l_m r_m);(l'_1 r'_1 \ldots l'_n r'_n)} = 2\,\left( E_{r_1}+ \ldots + E_{r_m} - E_{l_1} - \ldots - E_{l_m} \right) \cdot \left( \hat{v}_{j}^{(m,n)} \right)_{(l_1 r_1 \ldots l_m r_m);(l'_1 r'_1 \ldots l'_n r'_n)}
\end{align}
for products with non-zero matrices on the diagonal $(m,m)$. Here, the indices $(l_1 r_1 \ldots l_m r_m);(l'_1 r'_1 \ldots l'_n r'_n)$ precisely determine the scalar entries of the matrices. The analogous multiplications involving  $\Delta \hat{H}^{-}$ are evaluated by replacing $\Omega \rightarrow \delta\Omega$. In principle, these products are not restricted to the usage of $\hat{v}_{j}(t)$ and, thus, are  valid for a generic matrix of the dimension of $\hat{v}_{j}^{(m \pm 1,m)}$. It is sufficient to consider products where $\Delta \hat{H}^{\pm}$ is on the left, since in the end we need to evaluate the traces and all products can always be arranged such that $\Delta \hat{H}^{\pm}$ is on the left by relying on the cyclic invariance of the traces.

To lowest order in $\rho \, \Omega^2 \ll V_d$, we find that
\begin{equation}
\Omega^2 \,\sum_{\substack{l_k r_k}} \frac{ \left( \hat{v}_{j}^{(m,n)} \right)_{(l_1 r_1 \ldots l_k r_k \ldots l_m r_m);(l'_1 r'_1 \ldots l'_n r'_n)}}{\widetilde{E} + E_{l_k} - E_{r_k} + i\, \left( \delta + 2\,\gamma \right)} \approx 0 ,\quad
\Omega^2 \,\sum_{\substack{l'_k r'_k}} \frac{ \left( \hat{v}_{j}^{(m,n)} \right)_{(l_1 r_1 \ldots l_m r_m);(l'_1 r'_1 \ldots l'_k r'_k \ldots l'_n r'_n)}}{\widetilde{E} + E_{l'_k} - E_{r'_k} + i\, \left( \delta + 2\,\gamma \right)} \approx 0
\end{equation}
if the sum runs over indices occurring in $\hat{v}_{j}^{(m,m)}(E)$ with the abbreviation $\widetilde{E} = E - E_{l_1} - \ldots - E_{l_{k-1}} - E_{l_{k+1}}  - \ldots - E_{l_m} + E_{r_1} + \ldots +  E_{r_{k-1}} +  E_{r_{k+1}} + \ldots + E_{r_m}$. This result holds within the approximation where the energy levels of the detector's reservoirs are almost continuous so that the sum can be replaced by an integral, that is, $\sum_{\substack{l_k r_k}} \rightarrow \int \rho_l(E_{l_k}) \,\rho_r(E_{r_k})\,\textrm{d}E_{l_k}\,\textrm{d}E_{r_k}$. Additionally, the hopping amplitudes are assumed to depend weakly on the energy levels, hence, being constant, that is, $\Omega_{lr} \equiv \Omega$, and also the density matrices of the emitter and collector, respectively, are approximated to be constant, with $\rho_l(E_{l_k}) = \rho_l$ and $\rho_r(E_{r_k}) = \rho_r$. In order to understand that the integral vanishes, it is helpful to realize that the all entries of $\hat{v}_{j}^{(m,n)}$ essentially characterize higher-order retarded Green's functions which describe the averaged evolution of the density matrix. These Green's functions have poles in the lower half of the complex plane proportional to $[\widetilde{E} + E_{l_k} - E_{r_k} + i\, \delta]^{-1}$ which can be shown by iteratively solving the Laplace transformed differential equations~(\ref{eq:ODEforv_z}) for each $\hat{v}_{j}^{(m,n)}(E)$. Thus, a contour integral yields zero since the integrand decreases $\propto \, 1/ E^2$. On the contrary, if the sum does not include a summation over indices of $\hat{v}_{j}^{(m,n)}(E)$, it can be calculated within the same approximation as being
\begin{align}
\, & \Omega^2 \,\sum_{\substack{l_k r_k}} \frac{1}{\widetilde{E} + E_{l_k} - E_{r_k} + i\, \left( \delta + 2\,\gamma \right)} \approx - i \,\pi\,\rho_l\,\rho_r\,\Omega^2\,\left( V_d + \widetilde{E} \right) \approx -\,\frac{i}{2}\, D
\end{align}
to lowest order in $\rho \, \Omega^2 \ll V_d$ with $D = 2\,\pi\,\rho_l\,\rho_r\,\Omega^2\,V_d$.~\cite{Gurvitz:1997} Concerning this, the integral is separated into a singular part and the principal-value part. While the principal part redefines the energy levels, the singular parts lead to the presented result by relying on the Sokhatsky-Weierstrass theorem. Thus, one obtains
\begin{align}
 \hat{v}_c(E)  & \approx \frac{i}{4}\, D \cdot \left(2\, \textrm{tr} \left\{ \hat{v}_{x}^{(m-1,m-1)}(E)\right\} + 2\,i\,\frac{\delta\Omega}{\Omega}\, \textrm{tr} \left\{\hat{v}_{y}^{(m-1,m-1)}(E)\right\} \right)\notag\\
& \quad -  \textrm{tr} \left\{ \hat{v}_{x}^{(m-1,m)}(0) \cdot \left( (i\,E-\delta-2\,\gamma)\,\hat{{1}}^{(m,m)} +\frac{i}{2}\,\Delta \hat{H}^{+ \, (m,m)}  \right)^{-1} \cdot \, \Delta \hat{H}^{+ \, (m,m-1)} \right\},\\
 \hat{v}_e(E)  & \approx -\, \frac{i}{4}\, D \cdot \left(2\, \textrm{tr} \left\{ \hat{v}_{x}^{(m,m)}(E)\right\}   - 2\,i\,\frac{\delta\Omega}{\Omega}\,\textrm{tr} \left\{\hat{v}_{y}^{(m,m)}(E)\right\} \right)\notag\\
& \quad -  \textrm{tr} \left\{\Delta \hat{H}^{+ \, (m,m+1)}   \cdot \left( (i\,E-\delta-2\,\gamma)\,\hat{{1}}^{(m+1,m+1)} +\frac{i}{2}\,\Delta \hat{H}^{+ \, (m+1,m+1)}  \right)^{-1}\cdot \hat{v}_{x}^{(m+1,m)}(0) \vphantom{\frac{1}{1}} \right\},\\
 \hat{v}_i(E)  & \approx \frac{i}{4}\, D \cdot \left(-\,2\,\frac{\delta\Omega}{\Omega}\, \textrm{tr} \left\{ \hat{v}_{y}^{(m-1,m-1)}(E)\right\}  + 2\,i\,\frac{\delta\Omega^2}{\Omega^2}\,\textrm{tr} \left\{\hat{v}_{x}^{(m-1,m-1)}(E)\right\} \right)\notag\\
& \quad -  \textrm{tr} \left\{\hat{v}_{y}^{(m-1,m)}(0)  \cdot \left( (i\,E-\delta-2\,\gamma)\,\hat{{1}}^{(m,m)} +\frac{i}{2}\,\Delta \hat{H}^{+ \, (m,m)}  \right)^{-1} \cdot \, \Delta \hat{H}^{- \, (m,m-1)}  \right\} \\
\hat{v}_j(E) & \approx \frac{i}{4}\, D \cdot \left(2\,\frac{\delta\Omega}{\Omega}\, \textrm{tr} \left\{ \hat{v}_{y}^{(m,m)}(E)\right\} + 2\,i\,\frac{\delta\Omega^2}{\Omega^2}\,\textrm{tr} \left\{\hat{v}_{x}^{(m,m)}(E)\right\} \right)\notag\\
& \quad - \textrm{tr} \left\{\Delta \hat{H}^{- \, (m,m+1)}  \cdot \left( (i\,E-\delta-2\,\gamma)\,\hat{{1}}^{(m+1,m+1)} -\frac{i}{2}\,\Delta \hat{H}^{+ \, (m+1,m+1)}  \right)^{-1}\cdot\, \hat{v}_{y}^{(m+1,m)}(0) \vphantom{\frac{1}{1}} \right\} \,,
\end{align}
which couples ($m$,$m$) blocks to ($m-1$,$m-1$)  and ($m+1$,$m+1$) blocks, respectively. Note that it is trivial to take the inverses since the corresponding matrices are diagonal. Similar calculations eventuate in analogous expression for all the other terms by replacing $m$ with $m+1$ or $m-1$, respectively, that is, $ \hat{v}_f(E) = \hat{v}_e(E) (m \rightarrow m-1)$, $ \hat{v}_h(E) = \hat{v}_c(E)( m \rightarrow m+1)$, $ \hat{v}_k(E) = \hat{v}_j(E) ( m \rightarrow m-1)$ and $ \hat{v}_l(E) = \hat{v}_i(E)(m \rightarrow m+1)$. Inserting the achieved results for $\hat{v}_a(E)$ to $\hat{v}_l(E)$ into Eq.~(\ref{eq:letters}) and iteratively solving in $m$ yields the desired differential equation for $v^{(m)}_j(t) := \textrm{tr} \left\{\hat{v}_{j}^{(m,m)}(t)\right\}$ with $j=x$ after performing the re-Laplace transform, that is,
\begin{align}
\dot{v}_x^{(m)}(t) &= \left(-2\,\gamma -\, \frac{D}{2}\left( 1 + \frac{\delta\Omega^2}{\Omega^2}  \right) \right) \,v_x^{(m)}(t) + \frac{D}{2}\,\left( 1 - \frac{\delta\Omega^2}{\Omega^2} \right)\,v_x^{(m-1)}(t) - 2\,\epsilon_0\,v_y^{(m)}(t) \,.
\end{align}
Likewise calculations for each $v^{(m)}_j(t)$ finally end up in the rate equations
\begin{align}
\dot{v}_0^{(m)}(t) &= -\,\frac{D}{2}\,\left( 1 + \frac{\delta\Omega^2}{\Omega^2} \right)\,v_0^{(m)}(t) + \frac{D}{2}\,\left( 1 + \frac{\delta\Omega^2}{\Omega^2} \right)\,v_0^{(m-1)}(t) -D\,\frac{\delta\Omega}{\Omega}\,v_z^{(m)}(t) +D\,\frac{\delta\Omega}{\Omega}\,v_z^{(m-1)}(t) \\
\dot{v}_y^{(m)}(t) &= \left(-2\,\gamma -\, \frac{D}{2}\left( 1 + \frac{\delta\Omega^2}{\Omega^2}  \right) \right) \,v_y^{(m)}(t) + \frac{D}{2}\,\left( 1 - \frac{\delta\Omega^2}{\Omega^2} \right)\,v_y^{(m-1)}(t)  + 2\,\epsilon_0\,v_x^{(m)}(t) - 2\,\Delta\,v_z^{(m)}(t)\\
\dot{v}_z^{(m)}(t) &= -\,\frac{D}{2}\,\left( 1 + \frac{\delta\Omega^2}{\Omega^2} \right)\,v_z^{(m)}(t) + \frac{D}{2}\,\left( 1 + \frac{\delta\Omega^2}{\Omega^2} \right)\,v_z^{(m-1)}(t) -D\,\frac{\delta\Omega}{\Omega}\,v_0^{(m)}(t) +D\,\frac{\delta\Omega}{\Omega}\,v_0^{(m-1)}(t) + 2\,\Delta\,v_y^{(m)}(t)
\end{align}
for the special case of considering fluctuations in the $z$ direction. The general case is treated analogously.
%Since the solution of these rate equations have the form of propagating waves in the $m$-space, it is useful according to Ref.~\cite{Gurvitz:1998} to apply the Fourier transform defined in Eq.~\eqref{eq:fourier_transf_m} which leads to the differential equations

\section{Perturbative Solution of the Rate Equations}\label{sec:appendix_perturbative_solution}

In this appendix we present the perturbative solution for $\delta\Omega \ll \Omega$ of the rate equations~(\ref{eq:ODE}) which describe the averaged evolution of the density matrix with the initial conditions that $m=0$ electrons have penetrated to the right reservoir at $t=0$ so that the initial state of the system is described by $v_j^{(m)}(t=0) = \delta_{m,0} \cdot v_j^{(0)}(t=0)$, $j = 0,x,y,z$. Hereto, we introduce the Fourier transform
\begin{equation}\label{eq:fourier_transf_m}
 \widetilde{v}_j(q,t) = \sum_m v_j^{(m)}(t)\,e^{-\,imq} \,,
\end{equation}
with respect to $m$. Hence, $\langle \mathcal{R}^{(m)}(t)\rangle_{\xi}$ becomes $\langle \widetilde{\mathcal{R}}(q,t)\rangle_{\xi} = 1/2\cdot (\widetilde{\mathbf{v}}(q,t) \cdot \mathbf{\sigma})$ which eventuates in an expression of the conditional current in terms of $\widetilde{v}_j(q,t)$ by comparing to Eq.~(\ref{eq:weak_current}), that is,
\begin{equation}\label{eq:weak_number}
\left.\vphantom{\langle \Omega \rangle}\right._{\Pi_{\textrm{f}}}\langle I(\tau) \rangle_{\Psi(t_0)} = -\,\frac{i\,e}{\tau}\, \left. \partial q\, \ln \left[ \textrm{tr}\, \left\{  \Pi_f \cdot  \vphantom{\frac{1}{2}} (\widetilde{\mathbf{v}}(q,t) \cdot \mathbf{\sigma}) \right\} \right] \right|_{q=0}.
\end{equation}
If assuming that $\widetilde{\mathbf{v}}(q,t)$ is an analytic function of $q$ and $t$, it can perturbatively be expanded in order to describe the evolution of the system's density matrix within a weak measurement regime,
\begin{equation}\label{eq:perturbation_v_tilde}
\widetilde{\mathbf{v}}(q,t) = \sum_{n=0}^{\infty} \, \mathbf{u}_n(q,t)\, \left(\frac{\delta\Omega}{\Omega} \right)^{n} \,,
\end{equation}
where $n$ denotes the order of perturbation. Substituting this perturbation into Eq.~(\ref{eq:weak_number}) the zeroth-order contribution reads
\begin{equation}
\langle I(\tau) \rangle = -\,i\, \frac{e}{\tau} \,\left. \partial_q \ln \left[ \textrm{tr}\, \left\{  \vphantom{\int}   \Pi_f \cdot (   \mathbf{u}_0(q,t)  \cdot \mathbf{\sigma}) \right\} \right]  \right|_{q=0}
\end{equation}
and the first-order contribution, that is, the averaged WV for the current, is identified as
\begin{align}
\left.\vphantom{\langle \Omega \rangle}\right._{\Pi_{\textrm{f}}}\langle I(\tau) \rangle^{\textrm{weak}}_{\Psi(t_0)} &= -\,i\, \frac{e}{\tau} \left[   \textrm{tr}\, \left\{ \vphantom{\int}  \Pi_f \cdot  (\mathbf{u}_0(q=0,t) \cdot \mathbf{\sigma}) \right\}         \cdot              \textrm{tr}\, \left\{  \vphantom{\int} \Pi_f \cdot  (\left. \partial_q \mathbf{u}_1(q,t) \right|_{q=0} \cdot \mathbf{\sigma}) \right\} \right. \notag\\
& \quad - \, \left.      \textrm{tr}\, \left\{  \vphantom{\int} \Pi_f \cdot \mathbf{u}_1(q=0,t) \cdot \mathbf{\sigma}) \right\}         \cdot              \textrm{tr}\, \left\{ \vphantom{\int}  \Pi_f \cdot (\left. \partial_q \mathbf{u}_0(q,t) \right|_{q=0} \cdot \mathbf{\sigma}) \right\}      \right]  \cdot \left[  \textrm{tr}\, \left\{  \vphantom{\int}   \Pi_f \cdot (\mathbf{u}_0(q=0,t) \cdot \mathbf{\sigma}) \right\}   \right]^{-2}.
\end{align}
Thus, the WV is completely expressed in terms of the averaged density matrix. Further analysis now focuses on the evaluation of $\mathbf{u}_n(q,t)$ for $n = 0,1$, aiming at finding an illustrative expression for the WV. Inserting the power series of Eq.~\eqref{eq:perturbation_v_tilde} into the Fourier transformed rate equations~(\ref{eq:ODE}) leads to a set of differential equation for each $\mathbf{u}_n(q,t)$. The resulting equations for the lowest orders are
\begin{align}\label{eq:u0}
\frac{\partial}{\partial\,t}\,\mathbf{u}_0(q,t) &= \left[ G_0 + G_{\mathbf{k}}- \frac{D}{2}\,(1-e^{i\,q}) \cdot {1}_{(4)} \right] \cdot \mathbf{u}_0(q,t),\\
\frac{\partial}{\partial\,t}\,\mathbf{u}_1(q,t) &= \left[ G_0 + G_{\mathbf{k}}- \frac{D}{2}\,(1-e^{i\,q}) \cdot {1}_{(4)} \right] \cdot \mathbf{u}_1(q,t) - D\,(1-e^{i\,q}) \cdot G_{01} \cdot  \mathbf{u}_0(q,t) \label{eq:u1}\\
\frac{\partial}{\partial\,t}\,\mathbf{u}_2(q,t) &= \left[ G_0 + G_{\mathbf{k}}- \frac{D}{2}\,(1-e^{i\,q}) \cdot {1}_{(4)} \right] \cdot \mathbf{u}_2(q,t) -\, \frac{D}{2}  \cdot  G_{q} \cdot  \mathbf{u}_0(q,t)    \notag\\
&  - D\,(1-e^{i\,q}) \cdot G_{01} \cdot \exp \left[\left( G_0 + G_{\mathbf{k}}- \frac{D}{2}\,(1-e^{i\,q}) \cdot {1}_{(4)}  \right) t \right] \cdot \mathbf{u}_1(q,t)    \label{eq:u2}
\end{align}
Here, we have introduced
\begin{equation}
 G_{01} := \left(\begin{array}{cccc}
0 & 0 & 0 & 1\\                                                                                                                                                                                                       
0 & 0 & 0 & 0\\
0 & 0 & 0 & 0\\
1 & 0 & 0 & 0\\
\end{array} \right) \quad \textrm{,} \quad
 G_{10} := \left(\begin{array}{cccc}
0 & 0 & 0 & 0\\                                                                                                                                                                                                       
0 & 1 & 0 & 0\\
0 & 0 & 1 & 0\\
0 & 0 & 0 & 0\\
\end{array} \right) \quad \textrm{and} \quad
G_{q} := \left(\begin{array}{cccc}
1 - e^{i\,q} & 0 & 0 & 0\\                                                                                                                                                                                                       
0 & 1 + e^{i\,q} & 0 & 0\\
0 & 0 & 1 + e^{i\,q} & 0\\
0 & 0 & 0 & 1 - e^{i\,q}\\
\end{array} \right) \,.
\end{equation}
Higher orders in $\mathbf{u}_n(q,t)$ are not relevant for the expression for the WV. Pertinent for the expression of the WV are solutions for the special cases $\mathbf{u}_{n}(q=0,t)$ and $\partial_q \mathbf{u}_{n}(q,t) |_{q=0}$.

The initial conditions are equivalent to $\mathbf{u}_0(q,t=0) = (1 , v_x^{(0)}(t=0) , v_y^{(0)}(t=0) , v_z^{(0)}(t=0))$. Thus, $\mathbf{u}_0(q,t=0)$ does not depend on $q$ initially and $\mathbf{u}_n(q,t=0) \equiv 0 $ for $n \geq 1$. Furthermore, $\partial_q \mathbf{u}_n (q, t=0) |_{q=0} \equiv 0$ for all $n \in \mathbb{N}$ and $\textrm{tr}\{ \langle \rho(t) \rangle_{\xi} \} \equiv 1$ implies $(\mathbf{u}_0(q=0,t))_0 \equiv 1$ and $(\mathbf{u}_n(q=0,t))_0 \equiv 0$ at any time $t$. Additionally, $\partial_t \mathbf{u}_n (q=0, t) \equiv 0$ at any $t$ to keep $\textrm{tr}\{ \langle \rho(t) \rangle_{\xi} \}$ unchanged. %initial condition: at $t=0$, $m=0$ electrons have penetrated to the right reservoir; this implies that $\left( \vec{u}_0(q,t=0) \right)_j \equiv v_j^{(0)}(t=0)$ and $ \vec{u}_n(q,t=0)\equiv 0$ for $n \geq 1$

The perturbative solution is obtained iteratively with $ \mathbf{v}(t) = \exp\left[ \left( G_0 + G_{\mathbf{k}} \right) t \right] \cdot \mathbf{v}(0)$ where $\mathbf{v}(0) = (v_x^{(0)}(0), v_y^{(0)}(0), v_z^{(0)}(0))$. It reads 
\begin{align}
\mathbf{u}_0(q=0,t) &= \left( \begin{array}{c}
		      1 \\
		      \mathbf{v}(t)
		      \end{array} \right), \quad  
\mathbf{u}_1(q=0,t) = \left( \begin{array}{c}
		      0 \\
		      \mathbf{0}
		      \end{array} \right),\quad
\mathbf{u}_2(q=0,t) = \left( \begin{array}{c}
		      0 \\
		      D\,t \left( v_x(t) \, \mathbf{e}_x + v_y(t) \, \mathbf{e}_y \right)
		      \end{array} \right),\\
\partial_q \mathbf{u}_{0}(q,t) |_{q=0} &= i\,\frac{D}{2}\,t\, \left( \begin{array}{c}
					  1 \\
					  \mathbf{v}(t)
					  \end{array} \right),\quad
\partial_q \mathbf{u}_{2}(q,t) |_{q=0} = \left( \begin{array}{c}
					  0 \\
					  \mathbf{0}
					  \end{array} \right), \\
\partial_q \mathbf{u}_{1}(q,t) |_{q=0} &= i\,D\, \left( \begin{array}{c}
\mathbf{e}_z^{\,\,T} \cdot \left(  \left[ G_0+G_{\mathbf{k}} \right]^{-1} \cdot \exp\left[ \left( G_0+G_{\mathbf{k}} \right) t \right] - \left[ G_0+G_{\mathbf{k}} \right]^{-1}   \right) \cdot \mathbf{v}(t=0) \\
\left(  \left[ G_0+G_{\mathbf{k}} \right]^{-1} \cdot \exp\left[ \left( G_0+G_{\mathbf{k}} \right) t \right]      -\left[ G_0+G_{\mathbf{k}} \right]^{-1}       \right)   \cdot \mathbf{e}_z
\end{array} \right).
\end{align}
Hereafter we conclude that
\begin{equation} \label{eq:matrix_U0}
 \textrm{tr}\left\{ \vphantom{    \left. \partial_q \langle  U_0(q,\tau)  \rangle_{\xi} \right|_{q=0}       }       \Pi_f \cdot  (\mathbf{u}_0(q=0,t) \cdot \mathbf{\sigma})  \right\} =     \vphantom{\frac{1}{1}}        1+ \mathbf{v}(\tau) \cdot \mathbf{n}  , \quad \textrm{tr}\left\{ \vphantom{    \left. \partial_q \langle  U_0(q,\tau)  \rangle_{\xi} \right|_{q=0}       }       \Pi_f \cdot  (\mathbf{u}_1(q=0,t) \cdot \mathbf{\sigma})  \right\} = 0
\end{equation}
and
\begin{equation} \label{eq:matrix_U1} 
 -i\,\textrm{tr}\left\{ \Pi_f \cdot (\left. \partial_q \mathbf{u}_1(q,t)\right|_{q=0} \cdot \mathbf{\sigma}) \right\} =   -i \intop_{0}^{t} \frac{\partial}{\partial\,s}\,\textrm{tr}\left\{ \Pi_f \cdot (\left. \partial_q \mathbf{u}_1(q,s)\right|_{q=0} \cdot \mathbf{\sigma}) \right\}  \,\textrm{d}s    =D\,\intop_{0}^{\tau} v_z(s) + n_z(s) \,\textrm{d}s \,.
\end{equation}
Inserting  Eqs.~(\ref{eq:matrix_U1}) and~(\ref{eq:matrix_U0}) into the expression for the conditional value in Eq.~(\ref{eq:weak_number}) then yields the expression for the WV, that is, Eq.~(\ref{eq:same_result}), which is presented in the main text.

Similar, the second-order contribution to the WV is evaluated by noting that
\begin{equation}
- i \,\textrm{tr}\left\{ \Pi_f \cdot (\left. \partial_q \mathbf{u}_0(q,t)\right|_{q=0} \cdot \mathbf{\sigma}) \right\} = \frac{D\,\tau}{2} \left( \vphantom{\frac{1}{1}}   1+ \mathbf{v}(\tau) \cdot \mathbf{n} \right)
\end{equation}
and
\begin{equation}
 \textrm{tr}\left\{ \vphantom{    \left. \partial_q \langle  U_0(q,\tau)  \rangle_{\xi} \right|_{q=0}       }       \Pi_f \cdot  (\mathbf{u}_2(q=0,t) \cdot \mathbf{\sigma})  \right\} = \frac{D\,\tau}{2} \, \left(  \vphantom{\frac{1}{1}}     v_x(\tau)\,n_x + v_y(\tau)\,n_y \right)
\end{equation}
which eventuates in the expression
\begin{equation}\label{eq:2nd_order_result}
\left.\vphantom{\langle \Omega \rangle}\right._{\Pi_{\textrm{f}}}\langle I(\tau) \rangle^{\textrm{weak(2)}}_{\Psi(t_0)} = -\,\frac{e\,D^2\,\tau}{2} \cdot  \frac{v_x(\tau) \, n_x + v_y(\tau) \, n_y }{1 + \mathbf{n}\cdot \mathbf{v}(\tau)}.
\end{equation}
Note that the second-order contribution in Eq.~(\ref{eq:2nd_order_result}) has the same characteristics, that is, the same denominator, as the first-order term in Eq.~(\ref{eq:same_result}), which implies analogous conditions for divergencies or peculiar WVs.

\section{Analysis of the continuos detection}\label{sec:permutation}

Here, we derive Eq.~(\ref{eq:n_multiple_readouts}) of the main text for the case of a continuous detection. We start by assuming $\Delta t \ll \textrm{min} \{1/\omega
(\Omega/\delta\Omega), 1/\gamma (\Omega/\delta\Omega), 1/D
(\Omega/\delta\Omega)\}$. The evolution between two subsequent readouts is still exactly given by Eq.~(\ref{eq:ODE}) with the modified initial condition that precisely $n_k$ electrons have been read out at $t=t_k$ so that $\mathbf{v}^{(n)}(t_k) = \delta_{n,n_k} \cdot (1, v_x^{(n_k)}(t_k), v_y^{(n_k)}(t_k), v_z^{(n_k)}(t_k) )$. In order to solve these modified differential equations it is useful to introduce a vector $\mathbf{w}(t) = ( \mathbf{v}^{(0)}(t), \mathbf{v}^{(1)}(t), \ldots , \mathbf{v}^{(n)}(t), \ldots )$ where $n$ labels the number of transferred electrons so that Eq.~(\ref{eq:ODE}) reads
\begin{equation}\label{eq:ODE_for_big_w}
\frac{\textrm{d}}{\textrm{d}t} \mathbf{w}(t) = \left( M_1 +M_2 \right) \cdot \mathbf{w}(t) 
\end{equation}
with $M_1^{(m,n)} = (G_0 + G_{\mathbf{k}} - D/2 )\,\delta_{m,n} + \frac{D}{2} {1}_{(4)} \,\delta_{m,n+1}$ and $M_2^{(m,n)} = (G_1 + \,\frac{D}{2} \, 1_{(4)}) \,\delta_{m,n} + (G_2 - \frac{D}{2} \,  1_{(4)}) \,\delta_{m,n+1} $.
%\begin{equation}
% M_1 = \left(
%	\begin{array}{cccc}
%        G & 0 & 0 & \\
%        \frac{D}{2\hbar} {1} & G & 0 &  \\
%        0 & \frac{D}{2\hbar} {1} & G &  \\
%        0 & 0 & \frac{D}{2\hbar} {1} &  \\
%        & \vdots &  & \ddots
%        \end{array}
%       \right)
%\end{equation}
%with $G = G_0 + G_{\vec{k}} - D/2\hbar $ and
%\begin{equation}
% M_2 = \left(
%	\begin{array}{cccc}
%        F_1 & 0 & 0 & \\
%        F_2 & F_1 & 0 &  \\
%        0 & F_2 & F_1 &  \\
%        0 & 0 & F_2 &  \\
%        & \vdots &  & \ddots
%        \end{array}
%       \right).
%\end{equation}
This differential equation is solved trivially in the limit $\Delta t \ll 1/\omega (\Omega/\delta\Omega),
1/\gamma (\Omega/\delta\Omega), 1/D (\Omega/\delta\Omega)$ we
are interested in, by noting that  $M_1$ is a block-diagonalized matrix
in Jordan form, with the solution 
\begin{align}\label{eq:gen_solution}
\mathbf{v}^{(n_k)}(t) &= \exp\left[(G_0 + G_{\mathbf{k}} - \frac{D}{2}\,{1})(t - t_k)  \right] \cdot \left( {1}_{(4)} + (G_1 + \,\frac{D}{2} \, 1_{(4)}) \Delta t \right) \cdot \mathbf{v}^{(n_k)}(t_k) \,,\\
\mathbf{v}^{(n_k + m)}(t) &= f_m \cdot \exp\left[(G_0 + G_{\mathbf{k}} - \frac{D}{2 }\,{1})(t - t_k)  \right] 
\cdot \left((G_2 - \frac{D}{2} \,  1_{(4)}) \Delta t  + \frac{D}{2 } (t-t_k) ({1}_{(4)} + (G_1 + \,\frac{D}{2} \, 1_{(4)}) \Delta t) \right) \cdot \mathbf{v}^{(n_k)}(t_k) \,,\notag
\end{align}
where $f_m = \frac{1}{m!} \left( \frac{D}{2 } (t - t_k) \right)^{m-1}$ with $m \geq 1$. Within our approximation, in Eq.~(\ref{eq:gen_solution}) probability
 conservation is ensured by $\textrm{tr}\{
 \rho^{\textrm{(system)}}(t)\} = \sum_{n=0}^{\infty} v_0^{(n)}(t)
 \equiv 1$. Setting the reading time scales as the smallest in the
 problem, that is, $\Delta t \ll 1/D$, which
 corresponds to a continuous readout, the solution in Eq.~(\ref{eq:gen_solution}) becomes
\begin{align}
 \mathbf{v}^{(n_k)}(t) &= \exp\left[(G_0 + G_{\mathbf{k}})(t - t_k)  \right]  \cdot \left( {1}_{(4)} + G_1 (t-t_k) \right) \cdot \mathbf{v}^{(n_k)}(t_k) \notag\\
  \mathbf{v}^{(n_k+1)}(t) &= \exp\left[(G_0 + G_{\mathbf{k}})(t - t_k)  \right]   \cdot G_2 (t-t_k) \cdot \mathbf{v}^{(n_k)}(t_k) \notag\\
   \mathbf{v}^{(n_k + m)}(t) &= 0  \,\,,\,\,\, m \geq 2 \,.
\end{align}
This is the limit where at most one electron penetrates through the QPC between two subsequent readouts. 
The probability that exactly one electron will have accumulated in the collector within a readout period time is given by $ P(0;\Delta t) = \textrm{tr}\{A \cdot \mathbf{v}^{(n_k)}(t_k)\}_0$, while zero electrons penetrate with a probability $ P(1;\Delta t) = \textrm{tr}\{B \cdot \mathbf{v}^{(n_k)}(t_k)\}_0$, where the matrices $A$ and $B$ are given by  Eqs.~(\ref{eq:def_A_and_B}) and $\{\ldots \}_0$ denotes the zeroth component.
With these definitions the conditional number of transmitted electrons can be expressed as
\begin{equation}\label{eq:permutations}
  \left.\vphantom{\langle \Omega \rangle}\right._{\Pi_{\textrm{f}}}\langle n(\tau) \rangle_{\Psi(t_0)} =\\          \frac{                \mathbf{n} \cdot       \left(          \sum\limits_{m=0}^{N} m \sum\limits_{\textrm{perm}} \left[  A^{N-m} \cdot B^{m}  \right]_{\textrm{perm}}           \right)        \cdot            \mathbf{v}      }{         \mathbf{n} \cdot       \left(          \sum\limits_{m=0}^{N}  \sum\limits_{\textrm{perm}} \left[  A^{N-m} \cdot B^{m}  \right]_{\textrm{perm}}           \right)        \cdot            \mathbf{v}       }
\end{equation}
where $\mathbf{v}$ is evaluated at $t=0$. $N$ describes the total number of readouts and
$\sum_{\textrm{perm}}$ indicates the sum over all possible  orders of
$A$ and $B$ in the string of products $A^{N-m} B^m$. 
In the limit of $N \rightarrow \infty$, $\Delta t \rightarrow 0$ while
keeping $N \cdot \Delta t = \tau$ constant, the sum over all
permutations in Eq.~(\ref{eq:permutations}) can be analytically
evaluated.

Defining the auxiliary function $f(m,\Delta t, N) = \sum_{\textrm{perm}} \left[ A^{N-m} \cdot B^m  \right]_{\textrm{perm}} $, the numerator of Eq.~(\ref{eq:permutations}) is evaluated as follows
\begin{align}
\mathbf{n} &  \cdot \left( \sum_{m=0}^{N} m \cdot f(m,\Delta t, N)  \right) \cdot \mathbf{v} \notag\\
& = i \,\mathbf{n} \cdot \partial_q  \left[ \sum_{m=0}^{N}  f(m,\Delta t, N) \,  e^{-i m q} \right] \cdot \mathbf{v} \notag\\
& = i \,\mathbf{n} \cdot \partial_q  \left[ \sum_{m=0}^{N}   \sum_{\textrm{perm}} \left[ A^{N-m} \cdot (B \,  e^{-i q} )^m  \right]_{\textrm{perm}}  \right] \cdot \mathbf{v} \notag\\
& = i \,\mathbf{n} \cdot \partial_q  \left[ \left( 1 + (G_0 + G_{\mathbf{k}} + G_1 + G_2 \, e^{-iq})   \Delta t \right)^N  \right] \cdot \mathbf{v} \notag
\end{align}
where the last step is valid since $\sum_{m=0}^{N} \sum_{\textrm{perm}} \left[ A^{N-m} \cdot B^m  \right]_{\textrm{perm}} = (A+B)^N $ and due to the definitions in Eqs.~(\ref{eq:def_A_and_B}). In the limit $N\rightarrow \infty$, $\Delta t \rightarrow 0$, $N \cdot \Delta t = \tau$, this readily yields $\tau \, \mathbf{n} \cdot G_2 \cdot \exp((G_0 + G_{\mathbf{k}} + G_1 + G_2)\tau) \cdot \mathbf{v}(0)$. The denominator is treated analogously, which finally leads to Eq.~(\ref{eq:n_multiple_readouts}).

\end{widetext}

\end{appendices}

\bibliographystyle{apsrev4-1}
\bibliography{wv-decoherence_v2}

\end{document}